\documentclass[aps,prX,superscriptaddress,twocolumn]{revtex4}

\usepackage{graphicx}
\usepackage{subfigure}
\usepackage[utf8]{inputenc}
\usepackage{amsmath}
\usepackage{amssymb}
\usepackage{hyperref}
\usepackage{appendix}
\usepackage{color}
\usepackage[normalem]{ulem}

\newcommand{\nc}{\newcommand}
\nc{\bra}[1]{\langle #1|}
\nc{\ket}[1]{|#1\rangle}
\nc{\braket}[1]{\left\langle #1 \right\rangle}
\nc{\la}{\langle}
\nc{\ra}{\rangle}
\nc{\Tr}{\text{Tr}}
\nc{\tr}{\text{tr}}
\nc{\e}{\text{e}}
\nc{\Id}{\mathbb{1}}
\nc{\eps}{\varepsilon}
\nc{\bigO}{\mathcal{O}}
\nc{\dd}{\mathrm{d}}
\nc{\dN}{\mathrm{d}N}
\nc{\half}{\frac{1}{2}}
\nc{\app}[1]{Appendix~\ref{#1}}
\nc{\sdot}{\left(\cdot\right)}

\begin{document}
\renewcommand{\sectionautorefname}{Sect.}
\renewcommand{\subsectionautorefname}{Sect.}
\renewcommand{\figureautorefname}{Fig.}
\def\equationautorefname~#1\null{%
  Eq.~(#1)\null
}

\title{Quantum Predictive Filtering}

\author{Arne L. Grimsmo}\email{arne.loehre.grimsmo@usherbrooke.ca}
\affiliation{Département de Physique, Université de Sherbrooke, Sherbrooke, Québec, Canada J1K 2R1}
\author{Susanne Still}\email{sstill@hawaii.edu}
\affiliation{Department of Information and Computer Sciences and Department of Physics and Astronomy, University of Hawai`i at M{\=a}noa, Honolulu HI 96822, USA.}

\date{\today}

\begin{abstract}

How can relevant information be extracted from a quantum process? In many situations, only some part of the total information content produced by an information source is useful. 
Can one then find an efficient encoding, in the sense of retaining the largest fraction of relevant information? 
This paper offers one possible solution by giving a generalization of a classical method designed to retain as much relevant information as possible in a lossy data compression. A key feature of the method is to introduce a \emph{second} information source to define relevance. 
 We quantify the advantage a quantum encoding has over the best classical encoding in general, and we demonstrate using examples that a substantial quantum advantage is possible. We show analytically, however, that if the relevant information is purely classical, then a classical encoding is optimal.
 
\end{abstract}


\maketitle

\section{Introduction}
Predicting future outcomes based on past observations is a fundamental problem, not only for science and technology, but also for living organisms.
A central question is how much of the available data is {\em useful} for prediction.
This question is closely related to the challenge of finding quantitative measures of complexity. 
One such measure is the fraction of information about a dynamical process's past states that is needed to describe its future states \cite{shaw1984dripping,  grassberger1986toward, Crut88a, nemenman2000information, BNT01}.
Efficient use of information then boils down to storing only information that is relevant for prediction: if two representations of past events yield equally good predictions, the simpler one is typically preferable~\cite{Occam, jeffreys, chaitin, Vapnik98, Still2014}.  

We have shown recently that this type of information efficiency is directly related to efficient thermodynamic operation~\cite{StillSivakBellCrooks12, Grimsmo13c}, implying that considerations regarding predictive filtering might be of relevance to the design of power efficient small scale devices, man-made or natural. Since real environments are ultimately quantum, this immediately raises the question: 
is there an advantage gained by encoding the information in a quantum memory, rather than a classical memory? In this paper we investigate under what conditions a quantum memory can be more predictive
than a classical memory, and whether it can do so with higher efficiency.

This question is particularly enticing in a biological context. Living systems are masters at adapting to their environment, and predicting future events is key to their survival.
Efficient information use is found throughout the nervous system, and may constitute a building principle of biological computing machines, such as neurons and brains~\cite{Bialek12}. 
Filtering useful bits from ``nonpredictive clutter''~\cite{Bialek12} furthermore allows for thermodynamic efficiency~\cite{StillSivakBellCrooks12, Grimsmo13c}. 
Recent work has found that biophyiscal devices indeed evolved to use energy in a highly efficient manner \cite{Kinosita:2000, Cappello:2007, H2012}.

There is mounting evidence that quantum effects play an important role in the efficient operation of some microscopic biological systems~\cite{Engel07,Lee07,Panitchayangkoon10}. For example, the extremely high energetic efficiency of excitation transport in light-harvesting complexes in bacteria and plants may be using some form of quantum random walk~\cite{Mohseni08}. It has been shown that a delicate balance of noisy and coherent quantum processes is necessary to reach optimal efficiency~\cite{Mohseni08,Huelga13,Chin13}, indicating that quantum effects might be exploited for an evolutionary advantage.

\subsection{Extraction of relevant information from classical information sources}

Any continuous information source contains an infinite amount of information, but not all of this information is useful to the receiver at the other end of a communication channel. One would therefore like to delineate relevant from irrelevant information. Shannon \cite{Shannon48} addressed this problem by pointing out that the rate of an information source for a given quality of signal reproduction should be taken as the smallest amount of information required to specify the source, subject to a given constraint on average distortion. The distortion measure, which has to be chosen {\em ad hoc} by the practitioner, implicitly contains a notion of relevance \footnote{A classical text on rate-distortion theory is \cite{Berger71}.}.

Relevant information is treated explicitly in a similar framework, called the ``Information Bottleneck" method~\cite{Tishby99}. Given a data source $X$, and a relevant variable $Y$ that depends on $X$, the method finds an optimal encoding $X \to M$ of the data into a representation $M$, such that information about $Y$ is kept while irrelevant bits are filtered out. The method achieves this by solving a constrained optimization problem, minimizing mutual information $I[X,M]$, subject to a constraint on $I[M,Y]$. The quantity $I[X,M]$ is taken to measure the coding cost, in line with Shannon's work, and the quantity $I[M,Y]$ measures the amount of relevant information kept in the ``memory", $M$. It can not exceed the total relevant information the raw data contains, i.e., $I[M,Y] \leq I[X,Y]$, due to the data processing inequality. The trade-off between coding cost and relevant bits retained is controlled by a Lagrange multiplier which parameterizes a family of solutions. The resulting compression scheme finds sufficient statistics in the limit where all of the relevant information is kept, i.e., $I[M,Y] = I[X,Y]$. When applied to time series prediction, this limiting model can be used to quantify the complexity of a dynamical system, as mentioned above \footnote{For more details, see \cite{Still2014}, and references therein.}. 

\subsection{Organization of the paper}

In this paper we introduce a quantum generalization of the Information Bottleneck method in \sectionautorefname \ref{sect:problemdef}. The data source, $X$, is now quantum, and relevance is defined with respect to a second quantum data source, $Y$. The coding protocol we introduce here can be seen as a generalization of the protocol used in quantum rate-distortion coding \cite{Barnum00}.

Following the classical method, we find an encoding of the information in $X$ into a memory $M$, maximizing the information about $Y$ while discarding irrelevant information. To that end, we 
derive  
self-consistent equations that any optimal encoding must obey (\subsectionautorefname \ref{subsect:opt.encod} and Appendix A). These equations form the basis for an iterative algorithm (Appendix C) which allows us to illustrate the behaviour of optimal quantum encodings using a series of numerical examples (\sectionautorefname \ref{sect:examples}).

We show analytically that a quantum advantage is possible only for non-classical relevant information, \emph{i.e.}, only when the co-occurence statistics between $X$ and $Y$ can not be described by a classical probability distribution (Sect. \ref{sect:noquadvantage} and Appendix \ref{nqa}). 
A system operating in an environment where relevant features can be fully approximated by a classical model, therefore does not gain from encoding quantum information.

In general, however, using a quantum memory allows for storing more relevant information without having to increase the size of the memory to do so. Moreover, we demonstrate that it is possible to find encodings with more information about $Y$ than what is \emph{maximally} achievable for any classical memory. 
We specify the quantum advantage in \sectionautorefname \ref{sect:infoplane}, and analyze examples in \sectionautorefname \ref{sect:examples} to verify that there can be a significant quantum advantage. For the examples, we compute entanglement and quantum discord present in the optimal encodings to shed light on the role played by quantum correlations. 

We find that a quantum advantage is possible even when the map from $X$ to $Y$ breaks any entanglement with the memory (\autoref{sect:exB}).
We furthermore demonstrate in \autoref{sect:exC} that in a quantum process with redundant information, irrelevant features get filtered out: the numerical algorithm is able to pick out and purify only the relevant information.

\section{\label{sect:problemdef}Relevant Quantum Information Encoding}
\begin{figure}
\centering
\includegraphics{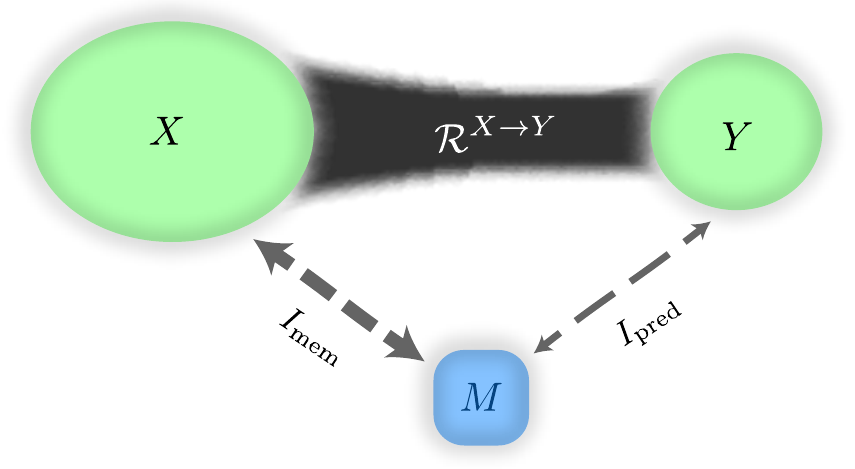}
\caption{\label{fig:qib}We consider a data source $X$, and define ``relevance'' by introducing a quantum channel, $\mathcal{R}^{X\to Y}$ to a second information source, $Y$. Information from $X$ is encoded in a memory $M$, and only information about $Y$ is deemed important. Relevant information is quantified by an information measure, $I_\text{pred}$, quantifying the correlations between $M$ and $Y$. We refer to $I_\text{pred}$ as ``predictive power.'' The correlations between $M$ and $X$, quantified by an information measure $I_\text{mem}$, we refer to as ``memory.'' The latter quantifies the total encoded information. An encoding is considered more efficient if it has greater $I_\text{pred}$ at the same $I_\text{mem}$.}
\end{figure}

Given a information source $X$, how can one introduce the notion that some information is more relevant than the rest? In conventional rate-distortion coding, the goal is to encode the information in $X$ with minimal distortion~\cite{Shannon48,Barnum00,Datta13a,Wilde13,Datta13b}. But in practice, it may be difficult to define an appropriate distortion measure. In general, we may want to encode only those aspects of the data that we deem important, and filter out the rest. This task is closely related to predictive inference, where we want to extract from $X$ exactly those features that are useful for making inferences about the outcome of some dynamical process that takes $X$ as its input. 

The Information Bottleneck method \cite{Tishby99} gives a simple solution to this problem: introduce a second information source $Y$, a ``relevant variable," which depends on $X$, and let this dependence \emph{define} relevance. Classically, the co-occurrance statistics of $X$ and $Y$ determine the available relevant information, which can be filtered out when the data are represented by a memory $M$. In a quantum mechanical generalization, two scenarios are possible: (i) think of $X$ and $Y$ as two quantum systems in a joint state $\rho_{XY}$, or (ii) think of $Y$ as output after sending $X$ through a quantum channel, $\rho_Y = \mathcal{R}(\rho_X)$. We here focus on the latter scenario, illustrated in \autoref{fig:qib}. Due to the causal relationship between $X$ and $Y$ in this scenario, we will refer to information about $X$ as ``memory'' and information about $Y$ extracted from $X$ as ``predictive information.'' The total amount of encoded predictive information  we call the encoding's ``predictive power.'' The map $\mathcal{R}^{X\to Y}$ defines relevance, and we refer to it as ``the relevance channel.''

  One cannot simultaneously send $X$ through two independent quantum channels, one with $Y$ as output and the other with $M$ as output. That is, we cannot both send $X$ through the relevance channel \emph{and} an independent encoding channel. Classically, this problem does not arise, and one can always make a copy of $X$, but quantum mechanically, no such physical process exists in general~\cite{Wooters98}. We therefore consider a protocol similar to that used in quantum rate distortion coding~\cite{Barnum00,Datta13a,Wilde13,Datta13b}: we take the input to the problem to be a purification of the state $\rho_X$, by introducing a second quantum system that acts as a reference, $R$. Information is then encoded in the memory by mapping $XR$ to $MR$. Subsequently, $MR$ is mapped to $MY$ via the relevance channel.

In the spirit of the classical Information Bottleneck method \cite{Tishby99}, we measure encoding cost and encoding quality using mutual information. Here we depart from the approach used in quantum rate distortion coding, as we do not use the usual qubit encoding rate to quantify cost, and the usual entanglement fidelity to quantify quality~\cite{Barnum00,Datta13a,Wilde13,Datta13b}.
Instead, we measure quality by predictive power, i.e., the relevant information quantifying the correlations between $M$ and $Y$. Memory cost is measured by mutual information between $M$ and $X$.

\subsection{Notation and definitions}

Hilbert spaces associated to quantum systems are denoted by $\mathcal{H}_A$, where the subscript is used to differentiate between systems. We assume that all Hilbert spaces have finite dimension, and let $d_A$ denote the dimension of system $A$, etc. The set of linear operators on $\mathcal{H}_A$, we denote $L(\mathcal{H}_A)$. We reserve the symbol $I_A \in L(\mathcal{H}_A)$ for the identity operator on $A$. A quantum state is a positive semi-definite operator $\rho_A \in L(\mathcal{H}_A)$, with unit trace. For a given state $\rho_{AB}$ defined on a composite system, $AB$, we use the convention that $\rho_A$ and $\rho_B$ denote the reduced states: $\rho_A = \tr_B \rho_{AB}$ and $\rho_B = \tr_A \rho_{AB}$, respectively.

Quantum channels, \emph{i.e.}, completely positive and trace preserving maps, are denoted by uppercase caligraphic letters, such as $\mathcal{E}^{A\to B}: L(\mathcal{H}_A)\to L(\mathcal{H}_B)$. The superscript is used to emphasize the input and output systems, but will be left out when a channel is applied to a state, \emph{e.g.}, $\mathcal{E}(\rho_A)$, to simplify notation. We use $\mathcal{I}_A$ to denote the identity channel on system $A$. A given quantum channel, $\mathcal{E}^{A\to B}$, can be extended to an ancilla system, $C$, by tensoring with the identity channel. Such an extension we denote by putting a hat on the same symbol as that used for the original channel: $\hat{\mathcal{E}}^{AC \to BC} = \mathcal{E}^{A\to B} \otimes \mathcal{I}_C$.

A purification of a state, $\rho_A$, is a pure state 
$\psi_{AR} = \ket{\psi_{AR}}\bra{\psi_{AR}}$, satisfying 
$\tr_{R} \,\psi_{AR} = \rho_A$. Here $R$ is a second quantum system with 
dimension at least as large as $A$. There are many possible 
purifications, but we here fix a choice with $R$ isomorphic to $A$:
\begin{align}\label{eq:purification}
  \ket{\psi_{AR}} = \sum_i \sqrt{p_i} \ket{i}\ket{i'},
\end{align}
where $\rho_A = \sum_i p_i \ket{i}\bra{i}$ is the spectral decomposition of 
$\rho_A$, and $\ket{i}$ and $\ket{i'}$ are orthonormal bases for the two 
systems $A$ and $R$, respectively.

The von-Neumann entropy of system $A$ in a state $\rho_A$ is defined as $S[A] = S(\rho_A) = -\tr[\rho_A \log \rho_A]$. The mutual information of two systems, $A$ and $B$, in a bipartite state, $\rho_{AB}$, is given in terms of von-Neumann entropies,
\begin{align}
  I[A:B] = I(\rho_{AB}) = S(\rho_A) + S(\rho_B) - S(\rho_{AB}).
\end{align}
Mutual information was given an operational meaning in~\cite{Groisman05}, where it was shown that it quantifies the smallest rate at which one must inject noise into the system for $A$ and $B$ to become uncorrelated, in the usual asymptotic limit. It can therefore be taken as a measure of the total correlations between the two systems.

We make use of the Choi-Jamiołkowski isomorphism to represent channels in 
terms of positive operators. More specifically, we use the framework 
developed in~\cite{Leifer13}, and introduce 
\emph{conditional quantum operators} to represent quantum channels. That is, 
any channel $\mathcal{E}^{A\to B}$ we can write uniquely in terms of a 
positive operator $\mathcal{E}_{B|A} \in \mathcal{L}(H_A\otimes H_B)$ satisfying $\tr_B \mathcal{E}_{B|A} = I_A$ and~\cite{Leifer13}
\begin{align}
\mathcal{E}(\rho_A) = \tr_A\left[\mathcal{E}_{B|A}^{T_A} \rho_A\right],
\end{align}
for all states $\rho_A$ on $A$. $T_A$ here denotes the partial transpose on $A$.

For \emph{given} initial state $\rho_A$ and channel $\mathcal{E}^{A\to B}$, we can define a bipartite quantum state on $AB$~\cite{Leifer13}:
\begin{align}
\rho_{AB} = \rho_A^{1/2} \mathcal{E}_{B|A} \rho_A^{1/2}.
\end{align}
Note that $\tr_B \rho_{AB} = \rho_A$ and $\tr_A \rho_{AB} = \rho_B = \mathcal{E}(\rho_A)$. Vice versa, for any \emph{given} bipartite state $\rho_{AB}$ we can define a channel through a conditional quantum operator 
\begin{align}
\mathcal{E}_{B|A} = \rho_A^{-1/2} \rho_{AB} \rho_A^{-1/2}. 
\end{align}
This framework allows us to switch back and forth between channels and states for representing the evolution of a given initial state.

Classical probability distributions we denote by $p(x)$ and $q(x)$, and conditional probability distributions by $p(y|x)$ and $q(y|x)$. A conditional probability distribution satisfies $\sum_y p(y|x) = 1$ for all $x$. A state on a bipartite system $XY$ that can be written
\begin{align}
\rho_{XY} = \sum_{x,y} p(x,y) \ket{x}\bra{x} \, \ket{y}\bra{y},
\end{align}
for a joint probability distribution $p(x,y)$ and orthonormal bases $\ket{x}$ for $\mathcal{H}_X$ and $\ket{y}$ for $\mathcal{H}_Y$, we call a \emph{classical state}, reflecting that $X$ and $Y$ are only classically correlated. A state that can be written
\begin{align}
\rho_{XY} = \sum_y p(y) \rho_{X|y} \ket{y}\bra{y},
\end{align}
where $\rho_{X|y}$ are arbitrary states of system $X$, we call \emph{quantum-classical}, reflecting that $X$ can be quantum while $Y$ is classical. Similarly a quantum channel, $\mathcal{E}^{X\to Y}$, is said to be \emph{classical} if it can be written in terms of a conditional quantum operator of the form
\begin{align}
\mathcal{E}_{Y|X} = \sum_{x,y} p(y|x) \ket{y}\bra{y} \, \ket{x}\bra{x},
\end{align}
where $p(y|x)$ is a conditional probability distribution. If this classical channel is applied to a state of $X$ that is diagonal in the same basis $\ket{x}$, $\rho_X = \sum_x p(x) \ket{x}\bra{x}$, we call this a \emph{classical process}. The output is in this case just $\rho_Y = \sum_y p(y) \ket{y}\bra{y}$, where $p(y) = \sum_x p(x,y) = \sum_x p(x)p(y|x)$.

\subsection{Relevant information optimization problem}

Denote by $\rho_X$ a state of a quantum system $X$, with purification $\psi_{XR}$. This state 
represents the data from which we wish to extract information. 
The reference system $R$ is sent through the \emph{relevance channel}, $\mathcal{R}^{R\to Y}$, with output system $Y$. We seek an optimal encoding of the information in $X$ to a memory, $M$, in terms of a quantum channel $\mathcal{E}^{X\to M}$, in the sense that we wish to retain as much information about $Y$ as possible, without storing any unnecessary data.

\begin{figure}
\centering
\includegraphics{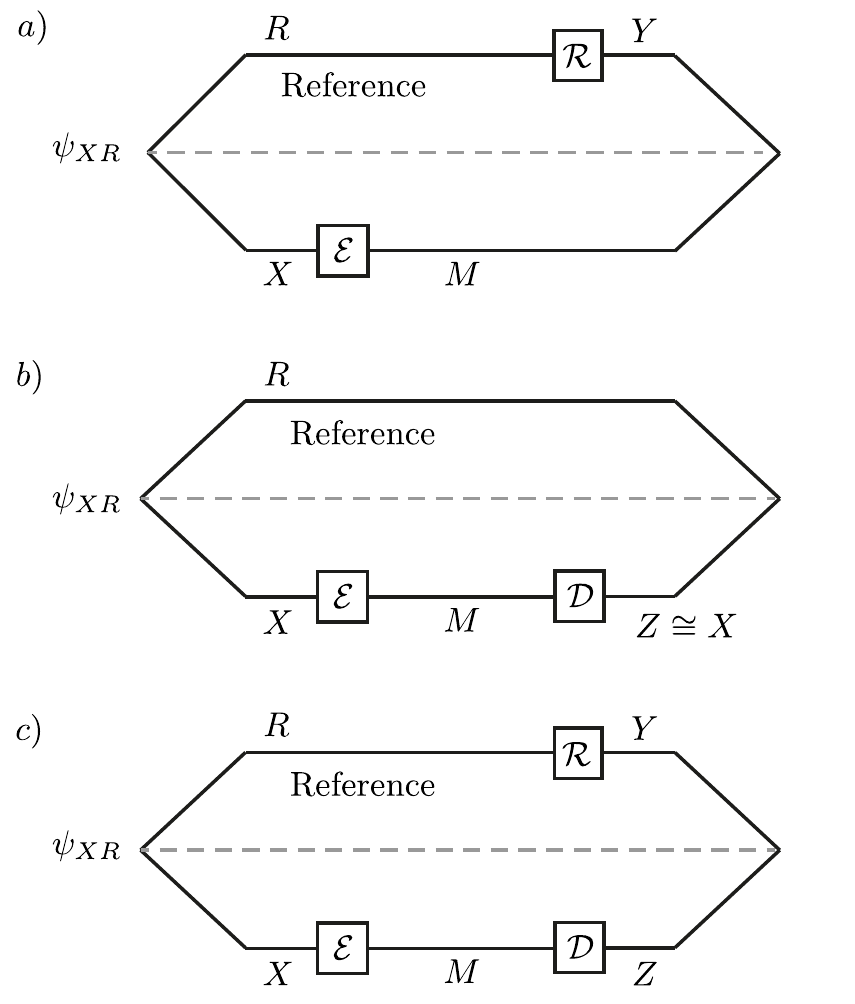}
\caption{\label{fig:rd}An illustration of the causal structure of three related coding protocols. In each case the input is a purification $\psi_{XR}$ of the initial data $\rho_X$. The protocols differ only by local maps applied to either the data, $X$, or the purification reference, $R$. $a)$~The ``relevance coding protocol'' considered in this paper. The data is encoded in a memory by an encoding map, $\mathcal{E}^{X\to M}$, while a ``relevance channel'' $\mathcal{R}^{R\to Y}$ is applied to the reference to define relevant information. $b)$~The conventional protocol for quantum rate distortion coding. No ``relevance channel'' is used, and all of the information in the initial data is considered equally important. A decoding map, $\mathcal{D}^{M\to Z}$, needs to be introduced to measure the fidelity of the encoded and original data. $c)$~A generalized rate distortion coding protocol, of which the two protocols in $a)$~and $b)$~are special cases.}
\end{figure}

The protocol we consider is similar to that used in quantum rate distortion coding~\cite{Barnum00,Datta13a,Wilde13,Datta13b}. Starting from the purification, $\psi_{XR}$, information is encoded in the memory by a map,
\begin{align}
\hat{\mathcal{E}}^{XR\to MR} = \mathcal{E}^{X\to M} \otimes \mathcal{I}_{R}.
\end{align}
This map is an extension of the local encoding map, $\mathcal{E}^{X\to M}$, to the reference system $R$. Information is now stored in the state 
\begin{align}
\rho_{MR} = \hat{\mathcal{E}}(\psi_{XR}).
\end{align}
Subsequently, the reference is sent through the relevance channel $\mathcal{R}^{R\to Y}$, which we extend to the memory system,
\begin{align}
\hat{\mathcal{R}}^{MR\to MY} = \mathcal{I}_M \otimes \mathcal{R}^{R\to Y},
\end{align}
so that information about $Y$ is finally encoded in the state 
\begin{align}\label{eq:relevance}
\rho_{MY} = \hat{\mathcal{R}}(\rho_{MR}).
\end{align}
The protocol is depicted in \autoref{fig:rd} $a)$.

We use mutual information to quantify both the information encoded about the initial data, which we refer to as \emph{memory}, 
\begin{align}
  I_\text{mem} = I_\text{mem}[\rho_{MR}] = I[M:R],
\end{align}
and the information available about the output on $Y$, which we refer to as \emph{predictive power},
\begin{align}
  I_\text{pred} = I_\text{pred}[\rho_{MR}] = I[M:Y].
\end{align}

Following the classical Information Bottleneck method, we wish to filter out the relevant information about $Y$. We do so by maximizing $I_\text{pred}$. But keeping more relevant information comes at the cost of increased memory, $I_{\rm mem}$. We define an \emph{optimal} encoding as one that maximizes $I_\text{pred}$, while simultaneously obeying a constraint on $I_\text{mem}$. Optimal encodings therefore must be solutions to the following optimization problem:
\begin{align}\label{eq:opt1}
\max_{\mathcal{E}^{X\to M}} I_\text{pred} \qquad \text{s.t.:} \qquad I_\text{mem} \le \text{const.}
\end{align}
Note that this problem is equivalent to minimizing $I_\text{mem}$ subject to a constraint on $I_\text{pred}$. 

Before we discuss solutions to this optimization problem, let us pause to consider the relationship to quantum rate distortion coding in more detail~\cite{Barnum00,Datta13a,Wilde13,Datta13b}. There, one also considers an information source $X$ in a state $\rho_X$, with purification $\psi_{XR}$, and seeks to find a minimal encoding of $X$ into a memory, $M$, given by a map $\mathcal{E}^{X\to M}$. The encoding is subject to a constraint quantifying the allowed degree of distortion. To measure the distortion, one introduces a decoding map, $\mathcal{D}^{M\to Z}$, where the system $Z$ is isomorphic to $X$. A distortion function measuring the fidelity of the decoded signal with the original, can then be introduced on the final output space of $Z$ and the reference system $R$. The protocol is depicted in \autoref{fig:rd} $b)$~\footnote{The encoding rate and the distortion measure is conventionally defined in a many-copy scenario: one considers $n$ copies of the information source, $X\to X^n$, $\rho_X \to \rho_X^{\otimes n}$, etc. The encoding and decoding channels are, respectively, maps from and back to the $n$-fold system space. The memory consists of $m$ qubits, and the coding rate is the number of memory qubits per copy of the source, $R=m/n$, as $n$ goes to infinity. The distortion measure introduced in~\cite{Barnum00} is based on average entanglement fidelity in the large $n$ limit, with high fidelity meaning low distortion. The optimization problem is to minimize the rate $R(D)$, over encoding and decoding maps, at a constraint on the allowed distortion, $D$.}.

A key difference between the protocol considered in this paper and quantum rate distortion coding is that in the latter, distortion is measured relative to the original data source, $X$, whereas we here measure it relative to a \emph{second} information source, $Y$. We seek to extract only the information from $X$ that is relevant for making an inference about $Y$. Conceptually, one could say that we measure average distortion by the negative mutual information, $-I_\text{pred} = -I[M:Y]$ (large $I_\text{pred}$ meaning low distortion)~\footnote{The choice of $I_\text{pred}$ and $I_\text{mem}$ as, respectively, quality and cost measures represents a direct quantum generalization of the classical Information Bottleneck method~\cite{Tishby99}. We note that $I_\text{mem}$ has an operational meaning as (two times) the qubit rate cost of simulating the channel, $\mathcal{E}^{X\to M}$, in the presence of arbitrary shared entanglement between the two parties $X$ and $M$~\cite{Devetak06,Abeyesinghe09}.}. 
Therefore, we do not need to introduce a decoding map to define distortion. More generally though, a decoding scheme can be introduced by introducing a map, $\mathcal{D}^{M\to Z}$, following $\mathcal{E}^{X\to M}$. This generalized protocol is depicted in \autoref{fig:rd} $c)$.

\subsection{Optimal encodings}
\label{subsect:opt.encod}
Finding the solutions to the optimization problem in \autoref{eq:opt1} is equivalent to solving
\begin{align}\label{eq:opt2}
  \begin{aligned}
  \rho_{MR}^\text{opt} = \arg \max_{\sigma_{MR}} \,&\Bigl( I\left(\hat{\mathcal{R}}(\sigma_{MR})\right)- \alpha I(\sigma_{MR}) \\
  &- \tr_{R}\left[\Lambda_{R} \sigma_{R}\right] \Bigr)
  \end{aligned}
\end{align}
(this is shown in \app{app:A}). Here $\alpha \ge 0$ and $\Lambda_{R}$ are Lagrange multipliers, where $\Lambda_{R}$ is a Hermitian operator on ${R}$. $\alpha$ is for the constraint on $I_\text{mem} = I[M:R]$, and $\Lambda_{R}$ for the constraint that $\sigma_{R} = \tr_M\,\sigma_{MR}$ must equal the given initial state $\rho_{R}$ (denoting the reduced state $\rho_R = \tr_X \psi_{XR}$, which is identical to the given initial state $\rho_X$, but defined on system $R$). Recall that $\hat{\mathcal{R}}^{R\to Y}$ is the relevance channel, see \autoref{eq:relevance}.

An optimal encoding $\mathcal{E}^{X\to M}_\text{opt}$ can be constructed from an optimal state found in \autoref{eq:opt2}, $\rho_{MR}^\text{opt}$, via the conditional quantum operator, 
\begin{align}\label{eq:opt:E_MgX}
\mathcal{E}_{M|X}^\text{opt} = \rho_{R}^{-1/2} \rho_{MR}^\text{opt} \rho_{R}^{-1/2}.
\end{align}
The optimal encoding map is represented as 
\begin{align}\label{eq:leiferspekkens}
\mathcal{E}_\text{opt}(\sigma_X) = \tr_X\left[(\mathcal{E}_{M|X}^\text{opt})^{T_X}\sigma_X\right].
\end{align}
Since $X$ and $R$ are isomorphic systems, we have introduced a slight abuse of notation by refering to the system $X$ on the left hand side of \autoref{eq:opt:E_MgX}, and $R$ on the right hand side. This simply serves to remind us that $\mathcal{E}_{M|X}$ is used to define an encoding map from $X$ to $M$.

As shown in \app{app:A}, an optimal encoding satisfies the following self-consistent equation
\begin{align}\label{eq:sol1a}
  \mathcal{E}_{M|X}^\text{opt} =&\, Z_R^{-1/2} \exp\Bigg( \log \rho_M^\text{opt} - \frac{1}{\alpha} H_{MR} \Bigg) Z_R^{-1/2},
\end{align}
where we have introduced
\begin{align}
  H_{MR} =& \hat{\mathcal{R}}^\dagger\left(\log\rho_{M}^\text{opt} + \log\rho_Y^\text{opt} - \log\rho_{MY}^\text{opt}\right),\label{eq:H_MR}\\
  Z_R =& \tr_M \left[ \exp\left( \log \rho_M^\text{opt} - \frac{1}{\alpha} H_{MR} \right) \right],\label{eq:Z_R}
\end{align}
and $\hat{\mathcal{R}}^{\dagger Y\to R}$ is the dual map of $\hat{\mathcal{R}}^{R\to Y}$~\footnote{The dual map is defined through $\tr_{MR} [\hat{\mathcal{R}}^\dagger(A_{MY})B_{MR}] = \tr_{MY} [A_{MY} \hat{\mathcal{R}}(B_{MR})]$, for all operators $B_{MR}$ on $MR$ and $A_{MY}$ on $MY$.}.
We have suppressed explicitly writing out identity operators in tensor products in Eqs. \eqref{eq:sol1a}--\eqref{eq:Z_R}, to keep the notation as simple as possible. 

\autoref{eq:sol1a} states an implicit relation that any optimal encoding must satisfy. In \autoref{sect:optimalproperties} we use this relation to derive several important properties of optimal encodings. Furthermore, \autoref{eq:sol1a} forms the basis for an iterative algorithm to explicitly find $\mathcal{E}_\text{opt}^{X\to M}$ for a given input state and relevance channel. In \autoref{sect:examples} we use this algorithm to find optimal encodings for a series of examples, and study the properties of these encodings. The iterative algorithm itself is discussed in \app{sect:algo}. \autoref{eq:sol1a} reduces to the well known classical result from~\cite{Tishby99} if the states are purely classically correlated (see \app{app:A:fullyclassical}).

\section{\label{sect:optimalproperties}Basic properties of optimal encodings}

\subsection{\label{sect:temp}Structure in the small and large $\alpha$ limits}

In the limit of $\alpha \rightarrow 0$, the second term in the exponential in \autoref{eq:sol1a} dominates. The operator thus has a form analogous to that of a thermal state, with $\alpha$ playing the role of temperature and $H_{MR}$ the role of Hamiltonian. As shown in \app{app:B}, the expectation value of $H_{MR}$, is 
the negative 
predictive power:
\begin{align}
  \braket{H_{MR}} = \tr[H_{MR}\rho_{MR}] = -I_\text{pred}(\rho_{MR}).
\end{align}
We therefore interpret the solution in this limit as a ``ground state'' that maximizes $I_\text{pred}$ by minimizing $\braket{H_{MR}}$. This is, of course, consistent with the fact that the constraint on $I_\text{mem}$ in the optimization problem vanishes as $\alpha \to 0$.

In the classical case, the solution found in this limit specifies a deterministic Hidden Markov Model which is a minimal representation with sufficient statistics \cite{Crut88a, Crut98d}. In the quantum case, this solution can be seen as a generalization of the concept of purification: while a purification is a solution that allows for retaining all of the relevant information, there are other, more efficient solutions.  A purification would follow from the choice $\mathcal{R}^{R\to Y} = \mathcal{I}_R$. The optimally predictive encoding, in contrast, is maximally correlated with the relevant information, as defined by the second information source, but uncorrelated with the irrelevant information.

In the large $\alpha$ limit, there is an infinite emphasis on compression. Then, the first term in the exponential dominates, and the term proportional to $1/\alpha$ can be neglected. \autoref{eq:sol1a} then implies that $\mathcal{E}_{M|X}^\text{opt} = \rho_M \otimes I_X$, where $\rho_M$ is arbitrary, and $I_\text{pred} = I_\text{mem} = 0$ in this limit.

\subsection{\label{sect:noquadvantage}No quantum advantage for classical processes}

Let $\ket{x}$ and $\ket{y}$ be orthonormal basis sets for the two systems $R$ and $Y$, respectively (recall that $R$ is isomorphic to $X$). Recall that the relevance map, $\mathcal{R}^{R\to Y}$, is said to be \emph{classical} if it can be written in terms of a conditional quantum operator of the form
\begin{align}\label{eq:R_classical}
\mathcal{R}_{Y|R} = \sum_{x,y} p(y|x) \ket{y}\bra{y} \, \ket{x}\bra{x},
\end{align}
where $p(y|x)$ is a conditional probability distribution. 
If, furthermore, the classical relevance channel in \autoref{eq:R_classical} is applied to an initial state $\rho_R$, that is diagonal in the basis $\ket{x}$, $\rho_R = \sum_x p(x) \ket{x}\bra{x}$, we call this a \emph{classical process}, since it can be described in terms of a stochastic map from a classical probability distribution to a classical probability distribution. 

An optimal encoding channel for classical processes is always of the \emph{classical} form
\begin{align}
  \mathcal{E}_{M|X}^\text{opt} = \sum_{m,x} p(m|x) \ket{m}\bra{m}\, \ket{x}\bra{x},
\end{align}
corresponding to an optimal state of the memory and reference 
\begin{align}\label{eq:CC_MX}
\rho_{MR}^\text{opt} = \sum_{m,x} p(x) p(m|x) \ket{m}\bra{m}\, \ket{x}\bra{x}
\end{align}
(this is shown in \app{app:B}). This means that {\em if} the relevance map deems only classical information about $X$ to be important, {\em then} quantum correlations are superfluous and should be filtered out in an optimal data representation.

By encoding information into non-orthogonal states of the memory, it is possible to reduce the entropy, $S(\rho_M)$, of the memory, without loss of predictive power. This was exploited in~\cite{Gu12} to construct a memory of smaller entropy than the best classical model, with no loss of information. However, we show in \app{app:B} that for a classical process, a quantum memory can not achieve higher predictive power as quantified by $I_\text{pred}$, than the best classical data representation, at the same $I_\text{mem}$. Hence, whenever the relevant data is purely classical there is no quantum advantage to predictive inference, as we have defined it here. 

This raises the question of whether the entropy of the memory is, in itself, a good measure of complexity~\cite{Gu12}. Our view is that, rather, the correlations between the memory and the source should be considered. Encoding classical information into non-orthogonal quantum states, does not require fewer input bits, even if the entropy of the memory alone can be lower. For classical states of the form \autoref{eq:CC_MX}, it is $I_\text{mem} = I[M:R]$ that correctly quantifies the encoding rate in classical bits~\cite{Shannon48,Tishby99}.

\section{\label{sect:infoplane}Quantifying quantum advantage: the information plane}

We do expect a quantum memory to generically perform better for non-classical processes. To quantify the quantum advantage, we compare the optimal solution of the quantum problem to the optimal solution when the 
memory 
is restricted to be classical. 

More precisely, we say that the memory is classical if the encoding maps leaves $MR$ in a \emph{classical-quantum} state $\sigma_{MR}$ of the form
\begin{align}\label{eq:MX_CQ}
  \sigma_{MR} = \sum_m p(m) \ket{m}\bra{m} \otimes \sigma_{R|m},
\end{align}
for a basis $\{\ket{m}\}$ on $M$, probability distribution $p(m)$, and arbitrary states $\sigma_{R|m}$ on $R$, with the constraint $\tr_M \sigma_{MR} = \rho_R$. The optimal classical encoding is thus defined to be the solution to the optimization problem \autoref{eq:opt2}, when the optimization is over the restricted set of states given by \autoref{eq:MX_CQ}. This restriction is discussed further in \app{app:A:MX_CQ}.

\begin{figure}
\centering
\includegraphics{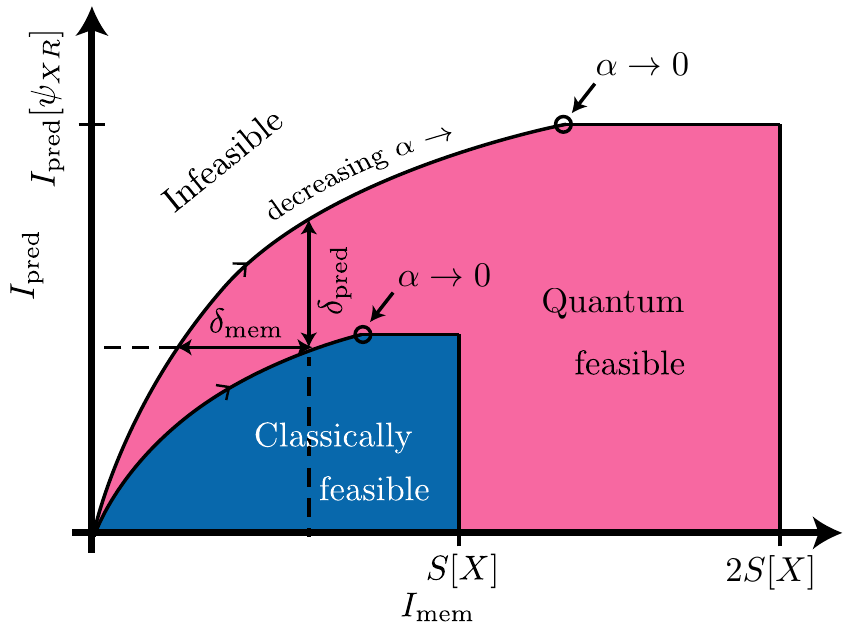}
\caption{\label{fig:infoplane}The information plane is spanned by $I_\text{mem}$ on the horizontal and $I_\text{pred}$ on the vertical axis. It can be divided into three regions: the infeasible, the quantum feasible and the classically feasible region. These regions can be mapped out using the quantum and classical upper bounds on $I_\text{mem}$ and the optimal curves found using the Quantum Information Bottleneck method. The optimal quantum curve generally lies above the optimal classical curve, and we quantify the gap by the two positive measures $\delta_\text{pred}$ and $\delta_\text{mem}$, as illustrated in the figure. The circles indicate the $\alpha \to 0$ limits of the optimal quantum and classical solutions.}
\end{figure}

We measure quantum advantage using the information plane~\cite{Tishby99}. The optimal values of $I_\text{mem}$ and $I_\text{pred}$ trace out a convex curve in the plane spanned by $I_\text{mem}$ and $I_\text{pred}$, as the Lagrange multiplier $\alpha$ is varied. We can find this curve numerically using the iterative algorithm presented in \app{sect:algo}. Whenever we need to distinguish between quantum and classical memories, we denote the values of $I_\text{mem(pred)}$ by $I_\text{mem(pred)}^\text{Q}$ and $I_\text{mem(pred)}^\text{C}$, evaluated for a quantum and a classical memory, respectively. The optimal curves for classical and quantum memories can in general be different, with the quantum curve lying above the classical. This generic situation is illustrated in \autoref{fig:infoplane}. 

We have the following general bounds on $I_\text{mem}$ and $I_\text{pred}$:
\begin{enumerate}
  \item The memory of a purification, $I_\text{mem}[\psi_{XR}] = 2S[X]$, upperbounds $I_\text{mem}$ for a any memory, quantum or classical, $I_\text{mem}^\text{C/Q} \leq I_\text{mem}[\psi_{XR}]$.
  \item The predictive power of a purification, $I_\text{pred}[\psi_{XR}]$, upperbounds $I_\text{pred}$ for any memory, quantum or classical, $I_\text{pred}^\text{C/Q} \leq I_\text{pred}[\psi_{XR}]$.
\item The entropy $S[X]$ of the initial data upperbounds $I_\text{mem}$ for a classical memory, $I_\text{mem}^\text{C} \leq S[X]$. 
\end{enumerate}
The last bound follows from writing $I_\text{mem} = S[R] - S[R|M]$, where $S[R|M] = S(\rho_{MR}) - S(\rho_M)$ is always greater or equal to zero for a classical memory of the form \autoref{eq:MX_CQ} (recall that $S[R] = S[X]$). A quantum memory can beat this bound by having negative values of the conditional entropy $S[R|M]$~\cite{Horodecki05,Rio11}.

The bounds introduced above, together with the curve traced out by the respective quantum and classical optimal encodings, can be used to define three regions of the information plane: (1)~the quantum feasible region containing all achievable points for quantum states satisfying the constraint $\tr_M \sigma_{MR} = \rho_R$; (2)~the classically feasible region, a subset thereof, which consists of the points achievable for states of the form of \autoref{eq:MX_CQ} that satisfy the same constraint; (3) the infeasible region containing those points corresponding to values of $I_\text{mem}$ and $I_\text{pred}$ that cannot be reached for any state satisfying the constraint. \autoref{fig:infoplane} illustrate these regions. The boundaries of the feasible regions in the informaiton plane allow us to read off the maximum predictive power at a given memory cost, as quantified by $I_\text{mem}$, for both quantum and classical memories.

We quantify the quantum advantage by comparing the optimal points achievable for a quantum memory to those of a classical memory. We introduce two measures: 
\begin{enumerate}
  \item $\delta_\text{pred}[I_\text{mem}] = I_\text{pred}^\text{Q}[I_\text{mem}] - I_\text{pred}^\text{C}[I_\text{mem}] \ge 0$, which we define to be the difference between the predictive power of the best quantum model and the best classical model, at a given $I_\text{mem}$.
\item $\delta_\text{mem}[I_\text{pred}] = I_\text{mem}^\text{C}[I_\text{pred}] - I_\text{mem}^\text{Q}[I_\text{pred}] \ge 0$, defined to be the difference in memory between the best classical and best quantum model at a given $I_\text{pred}$. 
\end{enumerate}
$\delta_\text{pred}$ is naturally only defined for $I_\text{mem} \le S[X]$, and $\delta_\text{mem}$ only for $I_\text{pred}$ smaller than the maximum achievable value for a classical memory. We emphasize that $\delta_\text{pred}$ and $\delta_\text{mem}$ quantifiy quantum advantage compared to \emph{any} classical memory of arbitrary dimension $d_M$. This is in contrast to quantum discord~\cite{Ollivier01,Henderson01} which in a sense compares a quantum memory to classical memories of the same dimension: Discord can be thought of as comparing a quantum encoding to an optimal classical encoding achievable through sending the quantum memory through a decoherence channel~\cite{Grimsmo13c}. 
Discord thus compares quantum bits to classical bits on a ``bit for bit'' basis, which is not necessarily a fair comparison in the present context.

\section{\label{sect:examples}Examples}

We consider three simple examples to illustrate how the quantum Information Bottleneck method can be used in practice, and to showcase some key features of optimal encodings. All optimal encodings are found numerically, using the algorithm in \app{sect:algo}.

The first example is a purely classical process. It serves to illustrate the fact that there is no quantum advantage for classical prediction problems, as indicated by our analytical results (\autoref{sect:noquadvantage}). We verify numerically that the optimal solutions for an arbitrary quantum memory are in agreement with those for a classical memory \cite{Still10} by tracing out the optimal curves in the information plane, showing that they overlap precisely.

The second example illustrates a process with a classical to quantum transition for the optimal encoding. A quantum bit is being sent through a phase damping channel. For compression values below the classical maximum, $I_\text{mem} \le S[X]$, the optimal classical memory performs just as well as the  quantum memory. But the predictive power so achieved does not exceed $65\%$ of the maximum possible predictive power. The optimal quantum memory can break through this classical barrier by taking on negative values of the quantum conditional entropy, $S[R|M]$. Thereby, the quantum memory can achieve {\em full} predictive power. This illustrates one of the main features of a quantum memory: it can achieve full predictive power on quantum processes where a classical memory can not. 

The third example shows a quantum process with redundant information. In this example, there is a quantum advantage for {\em any} value of retained memory. We quantify this quantum advantage. The redundant information (containing no utility for prediction) gets filtered out by the iterative algorithm that constructs the optimal encodings. 
Full predictive power is obtained at substantial compression: the maximum value of $I_\text{pred}[\psi_{XR}]$ is reached for a memory with almost three times less memory cost, $I_\text{mem} < I_\text{mem}[\psi_{XR}]$. We also find that this is achievable for a memory of smaller dimension than the initial data, $d_M < d_X$.

\subsection{Even process: a classical process}

Our first example illustrates that there is no quantum advantage for predicting a classical process. The example system is a hidden Markov process called the \emph{even process}~\cite{CruFeld03}. The predictive compressibility characteristics of this process were studied for a classical memory in~\cite{Still10}. We show here that the optimal quantum memory has identical features. 

 The even process outputs all binary strings consisting of an even number of 1s bounded by 0s, and associates a probability to any bit string by choosing either a 0 or a 1 with fair probability after having generated either a 0 or a pair of 1s. Consider the problem of predicting a future string of generated bits, based on having observed previous output strings. We chose the initial data to be the set of all bit strings, $x$, of length three, and we associate to each of them a probability, $p(x)$, equal to the frequency at which the even process generates the respective string. The relevance map is taken to be the map induced by the even process generating two new output bits from each respective previous sequence of three bits. To quantize this problem, we associate a set of pure orthogonal states $\ket{x}$, one for each history $x$, and take the state $\rho_X = \sum_x p(x) \ket{x}\bra{x}$ as initial data. There are eight possible bit strings of length three, so $d_X = 8$. Similarly, for the output we associate a set of pure orthogonal states $\ket{y}$, one for each possible two-bit output (so that $d_Y = 4$). The relevance map can then be defined by a conditional quantum operator $\mathcal{R}_{Y|R} = \sum_{x,y} p(y|x) \ket{y}\bra{y} \ket{x}\bra{x}$, where $p(y|x)$ is the stochastic map induced by the even process from the set of histories, $x$, to the set of futures, $y$.

\begin{figure}
\includegraphics{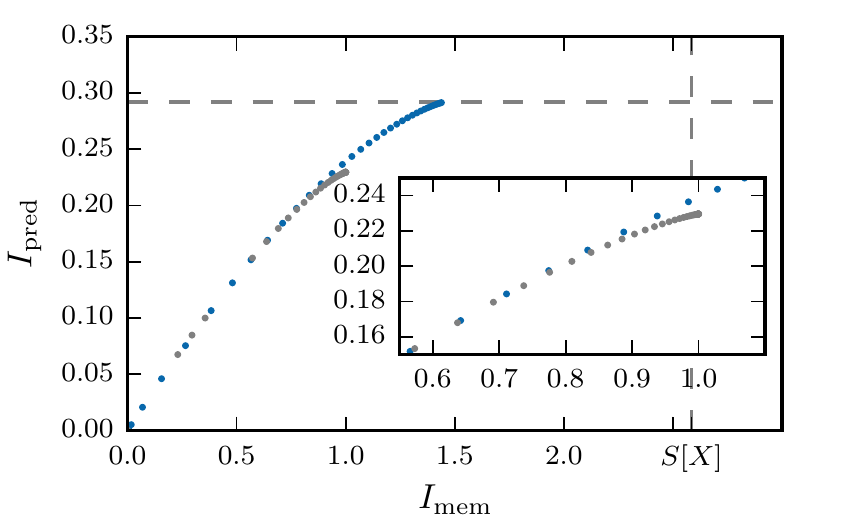}
\caption{\label{fig:evenprocess}Optimal curve in the information plane for the classical even process. Classical and quantum solutions coincide exactly. Dots show numerical results for a memory of dimension $d_M=3$ (blue), and $d_M=2$ (grey), for comparison. Inset: Zoom on region where the solutions for $d_M=3$ and $d_M=2$ bifurcate.}
\end{figure}

Both for a quantum memory, and for a classical memory, respectively, we compute the family of optimal encodings with the iterative algorithm and the deterministic annealing scheme described in \app{sect:algo}. The optimal values for the two cases exactly coincide, as shown in \autoref{fig:evenprocess}. The blue dots are the numerically found optimal values of $I_\text{mem}$ and $I_\text{pred}$, for {\em both} the quantum and the classical memory, as $\alpha$ is changed from a large value (lower left corner) to a small value. The maximal value of $I_\text{pred}$ is reached by a memory of dimension $d_M=3$. Our results are in agreement with~\cite{Still10}. 

The gray dots in \autoref{fig:evenprocess} shows the results for a memory of dimension $d_M=2$. This memory size is not large enough to reach the maximum value of $I_\text{pred}$, and the solution with $d_M=3$ bifurcates from that with $d_M=2$ as $\alpha$ is lowered, shown in the inset of \autoref{fig:evenprocess}. This illustrates that the algorithm can be used not only to find an optimal encoding at a fixed memory dimension, $d_M$, but also find the smallest possible $d_M$ reaching maximal predictive power. 

The maximal value of $I_\text{pred}$ is reached at a value of $I_\text{mem} \simeq 1.45$ bits, which is only $56\%$ of the classical entropy of the input data, $S[X] \simeq 2.6$ bits. Hence, significant compression of the input data is possible, without any loss of predictive power~\cite{Still10}. Note that the maximum quantum value is $2S[X] \simeq 5.2$ bits. This large value comes from the low degree of purity of the initial data, allowing a large degree of entanglement with the memory. Such entanglement is wasteful for predicting a classical process, and the algorithm correctly filters out this entanglement.

\subsection{\label{sect:exB}Causally incompressible quantum process}

This example serves to illustrate that there is a quantum advantage to predictive inference, even for quantum processes that are not causally compressible. 

The even process is an example of a (classical) causally compressible process ~\cite{StillCru07}. If a model can be constructed such that full predictive power is achieved at a memory less than the maximum set by the entropy of the data (in the classical case), then a process is fully causally compressible~\cite{StillCru07}. This definition is easily extended to quantum memories, with the maximum value for the memory now given by he memory of a purification, $I_\text{mem}[\psi_{XR}]$.

We look at a phase damping qubit channel. 
Phase damping describes loss of quantum information over time through decay of 
off-diagonal matrix elements for a quantum state. 
For a single qubit, the phase damping channel can be 
represented by the following operator-sum:
\begin{align}
  \mathcal{R}(\sigma_R) =& K_1 \sigma_R K_1^\dagger + K_2 \sigma_R K_2^\dagger,\\
  K_1 =& \left( \begin{array}{cc}
    1 & 0  \\
    0 & \sqrt{1-\lambda} \end{array} \right), \\
  K_2 =& \left( \begin{array}{cc}
    0 & 0  \\
    0 & \sqrt{\lambda} \end{array} \right),
\end{align}
where $\lambda = 1-\exp(-t/T_2)$ is a probability that grows with time,
$T_2$ being the qubit ``coherence time.''

As initial data we take the quantum state
\begin{align}\label{eq:ex2:rho_X}
  \rho_X = (1-p)\ket{+}\bra{+} + p \frac{I_2}{2},
\end{align}
where $\ket{+} = (\ket{0}+\ket{1})/\sqrt{2}$ and $I_2$ is the qubit identity 
operator. We choose parameters $\lambda = 0.5$ and $p = 0.3$. The qualitative nature of the results does not depend sensitively on the choice of parameters, but of course with a pure $\rho_X$ ($p=0$) we would necessarily have $I_\text{mem}[\psi_{XR}]=0$, and therefore no predictive compression would be possible.

\begin{figure}
\includegraphics{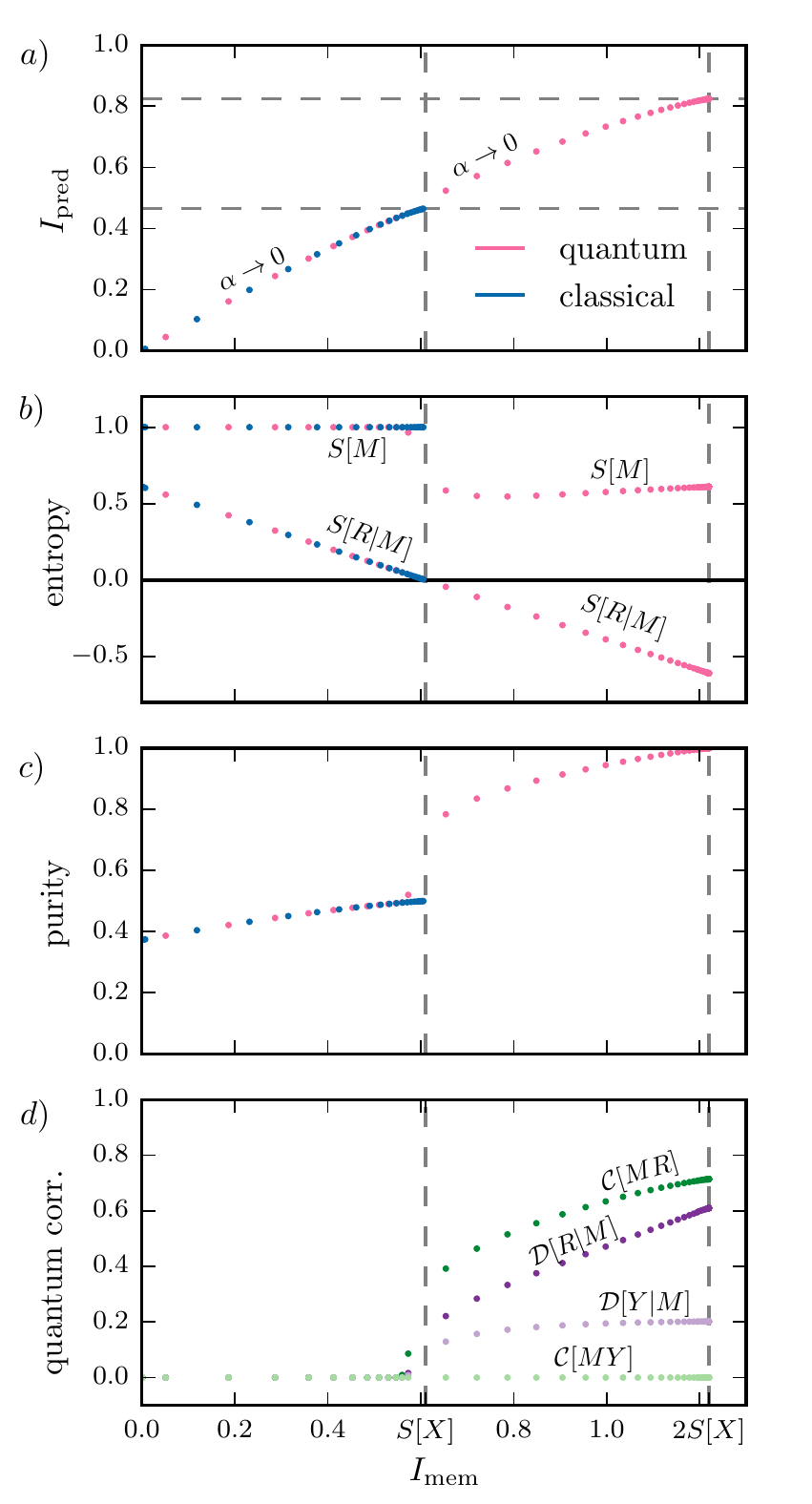}
\caption{\label{fig:phasedamp}Characteristics of optimal encodings for a qubit phase damping channel. (a-c): quantum solutions in pink, classical in blue. $a)$ Optimal solutions depicted in the information plane. $b)$ Conditional entropy $S[R|M]$ and memory entropy $S[M]$. $c)$ Purity of the optimal solution $\rho_{MR}^\text{opt}$. $d)$ Quantum correlations in the optimal solutions, quantified by concurrence (green) and quantum discord (purple), for the input (dark) and output (bright) systems, $MR$ and $MY$, respectively.}
\end{figure}

Numerical results for optimal classical and quantum memories are shown in 
\autoref{fig:phasedamp}, with blue and pink dots for the classical and quantum 
case, respectively. Panel $(a)$ shows the optimal curves in the 
information plane.
The predictive power of a classical memory is limited, it can not exceed $I_\text{pred} \simeq 0.46$ bits. This is only $65\%$ of the maximum possible predictive power of $I_\text{pred} \simeq 0.82$ bits. The quantum curve falls exactly on top of the classical curve, as long as $I_\text{mem} \le S[X] \simeq 0.61$. Beyond this value, we enter the classically infeasible region. The quantum memory is, however, able to break through this point. The quantum memory reaches the maximum possible predictive power, albeit requiring full complexity, $I_\text{mem} \simeq 1.2$ bits---twice the complexity of the best classical solution.

The point where the quantum memory goes beyond what is classically possible is associated with negative values of $S[R|M]$, and a sharp drop in the entropy of the memory, $S[M]$, as shown in panel $(b)$ of \autoref{fig:phasedamp}. This indicates a classical-quantum transition for the optimal encoding. This transition is further accompanied by a jump in purity at this point, which is shown in panel $(c)$ of \autoref{fig:phasedamp}. 
Note that in the limit of maximal predictive power, the optimal quantum solution is a purification of $\rho_R$, thus keeping all aspects of the original data and hence having full predictive power at full memory cost. 

The classical-quantum transition also comes along with a change in other quantities.
In panel $(d)$ of \autoref{fig:phasedamp}, we plot two different measures of 
quantum correlations for the optimal solutions. As a measure of entanglement 
between the the data qubit and the memory qubit we plot the 
concurrence~\cite{Wooters98}:
\begin{align}\label{eq:concurrence}
  \mathcal{C}[MR] = \mathcal{C}(\rho_{MR}) = \max(0,\lambda_1-\lambda_2-\lambda_3-\lambda_4)
\end{align}
where $\lambda_1,\dots,\lambda_4$ 
are the eigenvalues, in decreasing order, of the Hermitian matrix 
$\sqrt{\sqrt{\rho_{MR}}\tilde{\rho}_{MR}\sqrt{\rho_{MR}}}$. Here 
$\tilde{\rho}_{MR} = (\sigma_y \otimes \sigma_y)\rho^*_{MR}(\sigma_y\otimes\sigma_y)$, 
and $\sigma_y$ is the Pauli-$y$ matrix. We also plot the quantum discord 
\cite{Ollivier01,Henderson01} of the memory, 
\begin{align}\label{eq:discord}
  \begin{aligned}
    \mathcal{D}[R|M] = \mathcal{D}(\rho_{MR}) =& S(\rho_M) - S(\rho_{MR}) \\
  &+ \min_{\{P(m)\}} \sum_m p(m) S(\rho_{R|m}),
  \end{aligned}
\end{align}
where the minimization is over all sets of rank one projectors on $M$,
$\{P(m) = \ket{m}\bra{m} \otimes I_R\}$, such that $\sum_m P(m) = I_{MR}$, and 
$p(m) = \tr \left[P(m) \rho_{MR}\right]$, 
$\rho_{R|m} = \tr_M\left[ P(m) \rho_{MR} P(m) \right]/p(m)$. 

The discord is 
non-zero only for a non-classical memory, but is not a measure of non-local 
correlations like entanglement, as it can be created locally~\cite{Ciccarello12}.
Rather, it stems from encoding information into non-orthogonal states. In 
panel $(d)$ of \autoref{fig:phasedamp} we see that $\mathcal{C}[MR]$ and 
$\mathcal{D}[R|M]$ are zero for values of $I_\text{mem}$ below the 
classical-quantum transition, and non-zero above. In the $\alpha\to 0$ 
limit, the quantum memory shows a high degree of entanglement and discord. We 
also plot the concurrence and the discord for the output state $\rho_{MY}^\text{opt}$ 
(replacing $R$ with $Y$ in Eqs. \eqref{eq:concurrence} and 
\eqref{eq:discord}). Interestingly, we see that the relevance channel breaks the 
entanglement such that $\mathcal{C}[MY]$ is zero for all values of 
$I_\text{mem}$, while the discord, $\mathcal{D}[Y|M]$, is non-zero above the 
classical-quantum transition. 
This demonstrates a quantum predictive advantage even for an entanglement breaking process. The non-zero discord of the output state shows that there is a significant degree of ``quantumness'' of the predictive information encoded in the memory.

\subsection{\label{sect:exC}Quantum process with redundant information}

Our last example shows that redundant information is filtered out for fully causally compressible quantum processes. 
We take the initial data to be represented by two uncorrelated qubits. The relevance map acts on only one of them, while the other is discarded. The relevance channel is chosen in such a way that it can map orthogonal to non-orthogonal states, and is known to be able to create quantum discord~\cite{Ciccarello12}, even though it is a purely local map. In this example, there is a predictive advantage to having a quantum memory for \emph{any} complexity $I_\text{mem} >0$: the optimal quantum curve lies strictly above the optimal classical curve in the information plane.

The initial data conists of two qubits, $X=X_1X_2$, where the first is in the state given 
in \autoref{eq:ex2:rho_X} and the second is in a maximally mixed state:
\begin{align}\label{eq:redundantqubit}
  \rho_{X_1X_2} = \left[(1-p)\ket{+}\bra{+} + p \frac{I_2}{2}\right] \otimes \frac{I_2}{2}.
\end{align}
The relevance channel is taken to be an amplitude damping channel on the first qubit, while the second qubit is traced out,
\begin{align}
  \mathcal{R}(\sigma_{R_1R_2}) =& \tr_{R_2}\left[ K_1 \sigma_{R_1R_2} K_1^\dagger + K_2 \sigma_{R_1R_2} K_2^\dagger \right],\\
  K_1 =& \left( \begin{array}{cc}
    1 & 0  \\
    0 & \sqrt{1-\lambda} \end{array} \right) \otimes I_2, \\
  K_2 =& \left( \begin{array}{cc}
    0 & \sqrt{\lambda} \\
    0 & 0 \end{array} \right) \otimes I_2.
\end{align}
For the parameters we choose $\lambda=0.7$ and $p=0.3$. Again, the results do not depend sensitively on this choice of parameters. 

Clearly, the second qubit is irrelevant for making an inference about the first qubit. The choice of a maximally mixed state for the second qubit in \autoref{eq:redundantqubit} is made to have a high degree of mixedness in the initial data, which allows for large degree of correlation with a memory. However, information pertaining to the second qubit is redundant, and we use this example to show that the numerical algorithm filters out this information.

\begin{figure}
\includegraphics{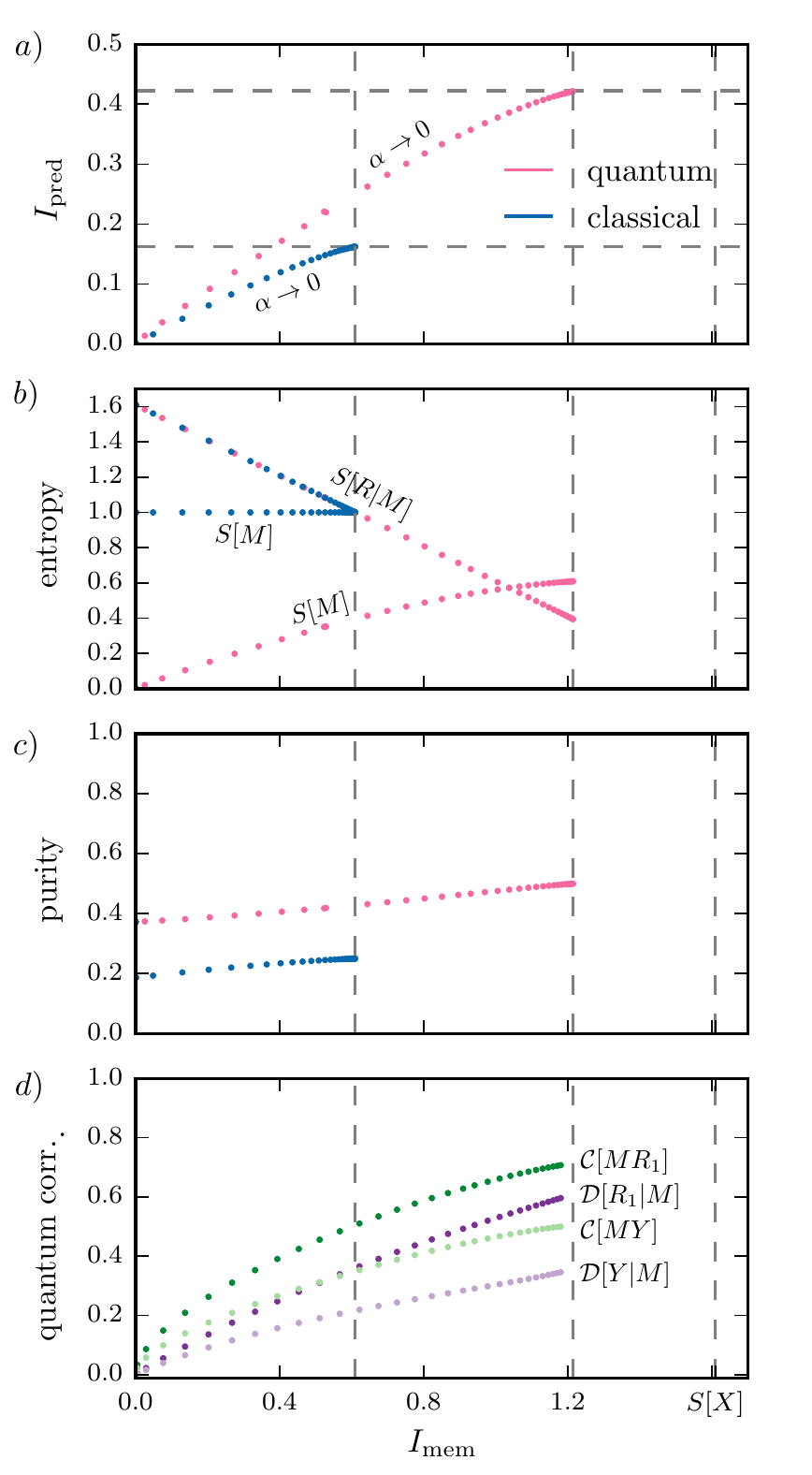}
\caption{\label{fig:ampdamp}Optimal encodings for an amplitude damping channel with redundant initial data. $a)$ The optimal quantum and classical curves in the information plane. $b)$ The conditional entropy $S[X|M]$ and the memory entropy $S[M]$. $c)$ The purity of the optimal solution $\rho_{MR}^\text{opt}$. $d)$ Quantum correlations in the optimal solutions, quantyfied by concurrence and quantum discord, for the input and output systems, $MR_1$ and $MY$, respectively, where $R_1$ refers to the first (relevant) data qubit.}
\end{figure}

In \autoref{fig:ampdamp} panel $(a)$ we show the optimal quantum and classical 
curves in the information plane, with the quantum curve lying strictly above 
the classical. We see that a very high degree of compression is possible. Importantly, the optimal quantum memory reaches maximum predictive power for $I_\text{mem} \simeq 1.2$
bits, well below
both the quantum and the classical maximal values 
of roughly $3.2$ bits and $1.6$ 
bits, respectively. 
The maximum is reached for 
a memory of dimension $d_M=2$, showing that the irrelevant quantum bit has 
been filtered out from the initial data. 

The classical memory is limited to a maximum predictive power of 
$I_\text{pred} \simeq 0.16$ bits, which is roughly 38\% of the maximum quantum 
value of $0.42$ bits.

To gain further insight into the origin of the quantum advantage, we plot in 
panel $(b)$ of \autoref{fig:ampdamp} the conditional entropy $S[R|M]$ and the 
memory entropy $S[M]$. 
In contrast to the previous example we now see that the quantum 
solution has a significantly lower memory entropy than the best classical 
solution, indicating that information is packed more densely into the quantum 
memory.

In \autoref{fig:ampdamp} panel $(c)$, we plot the purity, $\tr[\rho_{MR}^2]$, 
of the quantum 
and the classical solution. In contrast to the previous example, the optimal 
quantum solution in the $\alpha \to 0$ limit is no longer a purification of the 
initial data, due to the presence of the irrelevant data qubit in a highly 
mixed state. The maximum value of the purity 
for the quantum solution is found to be 0.5. This degree of mixedness comes solely from the 
entropy of the irrelevant qubit. Indeed, if we 
trace out this qubit, we find that the memory and the relevant data qubit is 
in a pure state: the optimal quantum solution in the 
$\alpha\to 0$ is found to be $\rho_{MR}^\text{opt}[\alpha\to 0] = \ket{\psi_{M R_1}}\bra{\psi_{M R_1}}\otimes I_2/2$, where $\ket{\psi_{M R_1}}$ refers 
to a purification of the first data qubit. This further illustrates how the 
irrelevant data qubit has been filtered out, and the numerical algorithm is able to pick out and purify only the relevant information.

The lower entropy of the quantum memory compared to the classical, illustrated 
in \autoref{fig:ampdamp} panel $(b)$, indicates that a predictive advantage is related to 
information encoded into non-orthogonal states. We therefore expect a non-zero 
quantum discord for the memory. Since we are not interested in quantum correlations
between the memory and the irrelevant second data qubit, we trace this qubit out
and consider the discord and the concurrence of the memory and the first, relevant,
data qubit, which is denoted $R_1$. The concurrence and discord are plotted in \autoref{fig:ampdamp} panel $(d)$. We show both the quantum 
correlations of the initial system, $MR_1$, and the output, $MY$. A high degree 
of both entanglement and discord in the initial and the final 
optimal state is seen to be necessary for an optimal encoding.

\begin{figure}
\includegraphics{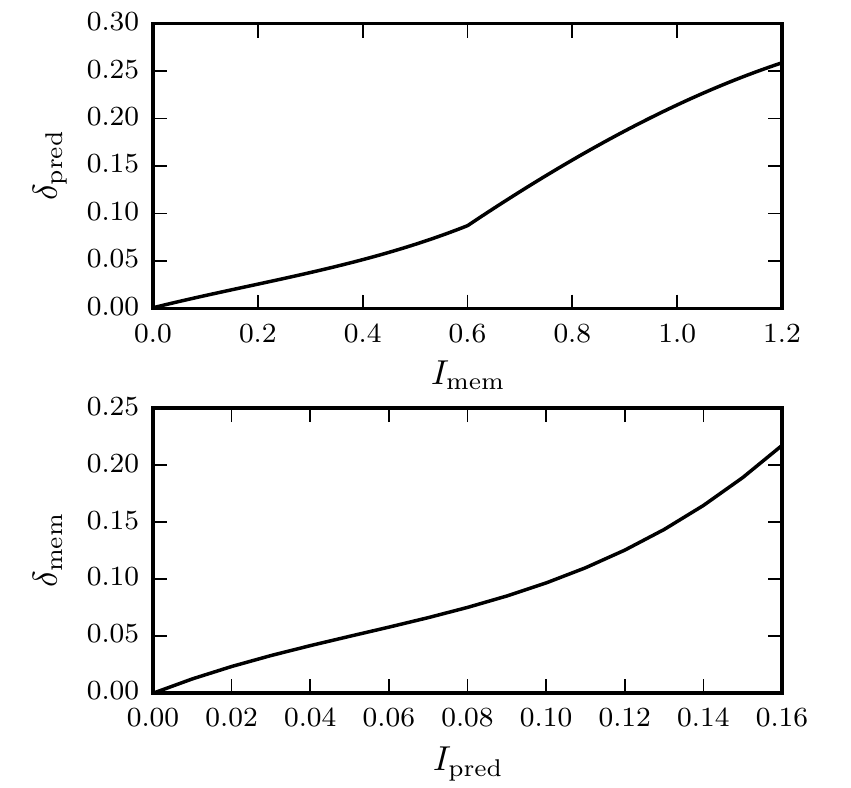}
\caption{\label{fig:ampdampquadvantage}Predictive quantum advantage for the amplitude damping channel, as quantified by $\delta_\text{pred}[I_\text{mem}]$ (top) and $\delta_\text{mem}[I_\text{pred}]$ (bottom).}
\end{figure}

Neither discord nor concurrence should be taken as measures of predictive quantum advantage. For this, we consider the measures $\delta_\text{pred}$ and $\delta_\text{mem}$ introduced in \autoref{sect:infoplane}. These quantities can essentially be read off from panel $(a)$ of \autoref{fig:ampdamp} but are displayed more clearly in \autoref{fig:ampdampquadvantage}. The values in this plot are calculated by interpolating the numerical data points displayed in \autoref{fig:ampdamp} panel $(a)$.

Quantum predictive advantage increases at the point where the classical limit is reached, \autoref{fig:ampdampquadvantage} (top). There is up to about 0.1 bits of quantum advantage in the classically feasible regime, and in the classically infeasible regime we find an additional quantum predictive advantage of as much as 0.26 bits. The quantum advantage to compression also becomes more pronounced as more predictive information is retained (bottom panel).

\section{\label{sect:conclusions}Conclusions}

We have introduced and analyzed a quantum generalization of the Information Bottleneck method, an approach to lossy compression that extracts relevant information. The coding protocol we introduced can be seen as a generalization of the protocol used in quantum rate-distortion coding. 

Both, the quality of the encoding and the coding cost are quantified in terms of mutual information. This allowed us to find analytically an implicit relation that any optimal encoding must satisfy. 

The approach we have taken allows 
for choosing
how much emphasis should be put on compression versus retention of relevant information. The limit where infinite weight is put on retaining relevant information is particularly interesting. In this limit, an encoding with full predictive information exists, often at significant compression. That is, the compressed data contains as much relevant information as does the initial uncompressed data \footnote{The trivial uncompressed encoding follows from $\mathcal{E}^{X\to M} = \mathcal{I}_X$ leaving the $MR$ in the purification state.}, but it can still be significantly less correlated with the initial data than a purification: the irrelevant information has been filtered out. For a quantum memory, this solution can thus be seen as a generalization of the concept of purification \footnote{A purification is achieved in this limit with the choice $\mathcal{R}=\mathcal{I}_R$ for the relevance channel)}. How much reduction can be achieved depends on the particular data source.

We have introduced a quantum generalization of the Information Bottleneck algorithm to find optimal encodings numerically. Examples were found exhibiting a substantial quantum predictive advantage. We studied a phase damping qubit channel, and found a classical-to-quantum transition: both classical and quantum encodings did equally well up to the point where the classical one reached its memory limit. The quantum encoding was able to capture all of the relevant information in the data (almost a factor two more than the best classical solution). Breaking through the classical memory limit was shown to be associated to a negative quantum conditional entropy. To achieve full predictive power, however, the memory had to be a purification of the initial data.

In contrast, a quantum process with redundant information can be compressed, one need only filter out the redundant bits. Using an example of two qubits, one of which was redundant, we showed that the numerical algorithm successfully picks out and purifies only the relevant qubit. In this example, there was a predictive advantage to having a quantum memory for any value of the memory kept.

We have shown analytically, and verified numerically that if the relevant information is purely classical, then there can be no quantum advantage. This result is interesting in the context of biological systems. It might be taken as an indication that an organism living in an environment where it suffices to make predictions about features that are fully approximated by a classical process, does not gain from encoding quantum information. This is intuitive, since we expect a possible quantum advantage only when the relevant features of the environment are intrinsically quantum.

\begin{acknowledgements}
SS is grateful for support from the Foundational Questions Institute (Grant No. FQXi-RFP3-1345). ALG is grateful for support by NSERC.
\end{acknowledgements}
\appendix

\section{\label{app:A}The self-consistent equations for an optimal solution}

In this appendix we show that a solution to the optimization problem, \autoref{eq:opt1}, fulfills \autoref{eq:sol1a} of the main text. We first derive the case for a general memory, and then consider the special case where the memory is restriced to be classical.

\subsection{Optimally predictive quantum memories}

We wish to solve the optimization problem
\begin{align}\label{eq:opt1_app}
\max_{\mathcal{E}^{X\to M}} I[M:Y] \quad \text{s.t.:} \quad I[M:R] \le \text{const.}
\end{align}
where
\begin{align}
&I[M:R] = S(\sigma_M) + S(\sigma_R) - S(\sigma_{MR}),\\
&I[M:Y] = S(\sigma_M) + S(\sigma_Y) - S(\sigma_{MY}),
\end{align}
and
\begin{align}
\sigma_{MR} =& \hat{\mathcal{E}}(\psi_{XR}),\\
\sigma_{MY} =& \hat{\mathcal{R}}(\sigma_{MR}).
\end{align}
Recall that $\rho_X$, with purification $\psi_{XR}$, and the relevance map $\mathcal{R}^{R\to Y}$, were assumed to be given. Also note that the reduced state on $R$, $\sigma_{R}=\tr_M\,\sigma_{MR}=\rho_R$, is identical to $\rho_X$, since $\hat{\mathcal{E}}^{XR\to MR}$ acts trivially on the reference $R$.

We first remark that optimizing over $\mathcal{E}^{X\to M}$ is equivalent to optimizing over all states, $\sigma_{MR}$, on the bipartite system $MR$, with the constraint
\begin{align}
    \tr_M\,\sigma_{MR} = \rho_R.
\end{align}
This is a straight forward consequence of the Choi-Jamiołkowski isomorphism, which gives one-to-one correspondence between channels and positive operators. More specifically, for a \emph{candidate} channel $\mathcal{E}^{X\to M}$, we introduce a representation in terms of a conditional quantum operator~\cite{Leifer13}:
\begin{align}\label{eq:app_A:channelrep}
\mathcal{E}(\sigma_X) = \tr_X[\mathcal{E}_{M|X}^{T_X}\sigma_X],
\end{align}
where $\mathcal{E}_{M|X}$ is a positive operator, satisfying $\tr_M \mathcal{E}_{M|X} = I_X$. $T_X$ denotes the partial transpose on $X$. We now introduce the following \emph{state} on $MR$ associated to $\mathcal{E}^{X\to M}$:
\begin{align}
    \sigma_{MR} = \rho_R^{1/2} \mathcal{E}_{M|X} \rho_R^{1/2}.
\end{align}
The slight abuse of notation here is based on the two systems $X$ and $R$ being isomorphic. For $\rho_R$ fixed this uniquely associates a state to the channel. On the other hand, given a state $\sigma_{MR}$ satisfying $\tr_M\,\sigma_{MR} = \rho_R$, we introduce the conditional quantum operator
\begin{align}\label{eq:app_A:condstate}
    \mathcal{E}_{M|X} =& \rho_R^{-1/2} \sigma_{MR} \rho_R^{-1/2},
\end{align}
which uniquely specifies a quantum channel, $\mathcal{E}^{X\to M}$, through \autoref{eq:app_A:channelrep}. For more details on the link between conditional quantum operators, as we define them here, and the Choi-Jamiołkowski isomorphism, we refer the reader to~\cite{Leifer13}.

This means that we can replace the maximization over $\mathcal{E}^{X\to M}$ by one over states, $\sigma_{MX}$, in \autoref{eq:opt1_app}. We next show explicitly that this is consistent, in the sense that $\hat{\mathcal{E}}(\psi_{XR}) = \rho_R^{1/2} \mathcal{E}_{M|X} \rho_R^{1/2}$, where $\mathcal{E}$ is given by \autoref{eq:app_A:channelrep}. First, we find for $\sigma_{MR}$:
\begin{align}
  \sigma_{MR} =& \mathcal{E}\otimes\mathcal{I}_{R}(\psi_{XR}) = \tr_{X}\left[\mathcal{E}_{M|X}^{T_X} \psi_{XR}\right]\\
  =& \sum_{ij} \sqrt{p_i p_j} \tr_X \left[\mathcal{E}_{M|X}^{T_X} \ket{i}\bra{j} \right] \ket{i'}\bra{j'} \\
  =& \sum_{ij} \sqrt{p_i p_j} \braket{j|\mathcal{E}_{M|X}^{T_X}|i} \ket{i'}\bra{j'}\\
  =& \sum_{ij} \sqrt{p_i p_j} \braket{i|\mathcal{E}_{M|X}|j} \ket{i'}\bra{j'}.\label{eq:sigma_MX1}
\end{align}
In the first line we introduced the representation of the map in terms of the conditional state. In the second line, we used \autoref{eq:purification} to write $\psi_{XR}$ in terms of basis states $\ket{i}$ and $\ket{i'}$ on the two respective systems, $X$ and $R$. In the third line we performed the partial trace, and in the last line the partial transpose. On the other hand, we also have that
\begin{align}
  \sigma_{MR} =& \rho_R^{1/2} \mathcal{E}_{M|X} \rho_R^{1/2}\\
  =& \sum_i \sqrt{p_i} \ket{i}\bra{i} \mathcal{E}_{M|X} \sum_j \sqrt{p_j} \ket{j}\bra{j}\\
  =& \sum_{ij} \sqrt{p_i p_j} \braket{i|\mathcal{E}_{M|X}|j} \ket{i}\bra{j}.\label{eq:sigma_MX2}
\end{align}
where we simply inserted the spectral decomposition $\rho_R = \sum_i p_i \ket{i}\bra{i}$. Comparing \autoref{eq:sigma_MX1} and \autoref{eq:sigma_MX2}, we see that the expressions are identical.

In summary, we now consider the optimization problem
\begin{subequations}\label{eq:opt2_app}
\begin{align}
  &\max_{\sigma_{MR}} I\left(\hat{\mathcal{R}}(\sigma_{MR})\right)\\
  {\rm s.t.:}\;\;\; &I(\sigma_{MR}) \le \text{const.}\\
  &\sigma_R = \tr_M [\sigma_{MR}] = \rho_R.
\end{align}
\end{subequations}
We solve the constrained optimization problem in \autoref{eq:opt2_app} by introducing Lagrange multipliers. That is we seek to maximize a Lagrangian
\begin{align}
  \max_{\sigma_{MR}} L(\sigma_{MR}),
\end{align}
defined by
\begin{align}\label{eq:opt3_app}
  \begin{aligned}
  L(\sigma_{MR}) =& I\left(\hat{\mathcal{R}}(\sigma_{MR})\right) - \alpha I(\sigma_{MR}) \\
                  &- \tr_{R}\left[\Lambda_{R} \sigma_{R}\right].
  \end{aligned}
\end{align}
Here $\alpha \ge 0$ and $\Lambda_{R}$ are the Lagrange multipliers, where $\Lambda_{R}$ is a Hermitian operator on ${R}$. $\alpha$ is for the constraint on $I(\sigma_{MR})$, and $\Lambda_{R}$ for the constraint that $\sigma_{R} = \tr_M\,\sigma_{MR} = \rho_{R}$ is the given initial state. To clarify the latter constraint, it is useful to expand $\rho_R$ in some orthonormal basis of Hermitian operators $H_{kl}$ on $\mathcal{H}_R$, $\rho_R = \sum_{k,l} x_{kl} H_{kl}$, $x_{kl} \in \mathbb{R}$. The constraint that $\tr_M\,\sigma_{MR} = \rho_{R}$ can then be stated as $\tr_{R}[H_{kl}\,\tr_M\,\sigma_{MR}] =  x_{kl}$, for all $k,l$. Hence we can write 
\begin{align}
  \Lambda_{R} = \sum_{kl} \lambda_{kl} H_{kl},
\end{align}
with $d_X^2$ real Lagrange multipliers $\lambda_{kl}$. 

We proceed by considering a small variation in $\sigma_{MR}$: $\sigma_{MR} \to \sigma_{MR} + \delta_{MR}$, for traceless and Hermitian $\delta_{MR}$. The functional derivative, $\delta L / \delta_{MR}$, is then defined through
\begin{align}
  \delta L = \tr \left[ \frac{\delta L}{\delta_{MR}} \delta_{MR} \right],
\end{align}
where $\delta L$ is the variation in $L$ to first order in $\delta_{MR}$. 
In particular, we have the following functional derivatives:
\begin{align}
  &\frac{\delta S(\sigma_M)}{\delta_{MR}} = - \log \sigma_M \otimes I_R - I_{MR}, \\
  &\frac{\delta S(\sigma_R)}{\delta_{MR}} = - I_M \otimes \log \sigma_R - I_{MR}, \\
  &\frac{\delta S(\sigma_{MR})}{\delta_{MR}} = - \log \sigma_{MR} - I_{MR}, \\
  &\frac{\delta S(\sigma_Y)}{\delta_{MR}} = - \hat{\mathcal{R}}^\dagger \left( I_M \otimes \log \sigma_Y + I_{MY} \right), \\
  &\frac{\delta S(\sigma_{MY})}{\delta_{MR}} = - \hat{\mathcal{R}}^\dagger \left( \log \sigma_{MY} + I_{MY} \right).
\end{align}
Note that since $\hat{\mathcal{R}}^{MR\to MY}$ is trace-preserving, the dual map, $\hat{\mathcal{R}}^{\dagger MY\to MR}$, is unital, \emph{i.e.}, $\hat{\mathcal{R}}(I_{MY}) = I_{MR}$. We can now write down the variational condition for the optimum, $\delta L / \delta_{MR} = 0$,
\begin{align}
  \frac{\delta L(\sigma_{MR})}{\delta_{MR}} =& -\log \sigma_M + \hat{\mathcal{R}}^\dagger \left(\log \sigma_{MY} - \log \sigma_{Y}\right) \nonumber \\
  &+ \alpha \left( \log\sigma_M + \log\sigma_R - \log\sigma_{MR} \right) \\
  &- \tilde{\Lambda}_{R}= 0, \nonumber
\end{align}
where we have absorbed an operator proportional to $I_{MR}$ into the Lagrange multiplier $\tilde{\Lambda}_R$, and suppressed the identity operators in tensor products, to keep the notation as clean as possible.
Rearranging this, we have
\begin{align}\label{eq:app_A:sol1}
  \sigma_{MR} =& \exp\Bigg( \log \sigma_M + \log \sigma_R - \frac{1}{\alpha} H_{MR} - \tilde{\Lambda}_R \Bigg),
\end{align}
where $H_{MR}$ is defined in \autoref{eq:H_MR}.

In \autoref{eq:app_A:sol1}, the Hermitian operator $\tilde{\Lambda}_R$ must be chosen such that $\tr_M \sigma_{MR} = \rho_R$. Since this involves varying $d_X^2$ real Lagrange multipliers, finding such a $\tilde{\Lambda}_R$ is costly. A better approach is to rewrite the exponential 
\begin{align*}
  &\exp(\log \sigma_M + \log \sigma_R - H_{MR}/\alpha - \tilde{\lambda}_R)  \\
  &= C_R \exp(\log \sigma_M - H_{MR}/\alpha) C_R^\dagger,
\end{align*}
where $C_R$ is an operator acting non-trivially on system $R$ only. To see that this is possible, consider the Zassenhaus formula for the expansion of an exponential~\cite{Casas12},
\begin{align}
  \e^{X+Y} = \e^{X}\e^{Y} \prod_{n=2}^\infty \e^{C_n(X,Y)},
\end{align}
where $C_n(X,Y)$ is a polynomial of degree $n$ in $X$ and $Y$ that can be written in terms of nested commutators of the two operators (with $[X,Y]$ as the innermost commutator). For example, the three first $C_n(X,Y)$ are 
\begin{align*}
  C_2(X,Y) =& -\frac{1}{2} [X,Y],\\
  C_3(X,Y) =& \frac{1}{3}[Y,[X,Y]] + \frac{1}{6}[X,[X,Y]],\\
  C_4(X,Y) =& \frac{1}{24}4[[[X,Y],X],X] + \frac{1}{8}[[[X,Y],X],Y] \\
  &+ \frac{1}{8}[[[X,Y],Y],Y].
\end{align*}
To apply this to \autoref{eq:app_A:sol1}, we identify $X = \log \sigma_M - H_{MR}/\alpha$ and $Y = \log \sigma_R - \tilde{\Lambda}_R$, so that $Y$ is an operator on system $R$ only. It is then clear that the commutator $[X,Y]$ acts non-trivially on system $R$ only, as well. We can therefore write 
\begin{align*}
  \e^{X + Y} =& \e^{X+Y/2}\e^{Y/2} \prod_{n=2}^\infty \e^{C_n(X+Y/2,Y/2)}\nonumber  \\
  =& \left(\prod_{n=2}^\infty \e^{C_n(X+Y/2,Y/2)}\right)^\dagger \e^{Y/2}\e^{X+Y/2}  \nonumber \\
  =& \left(\prod_{n=2}^\infty \e^{C_n(X,Y/2)}\right)^\dagger \e^{Y/2}\e^{X+Y/2}  \\
  =& \left(\prod_{n=2}^\infty \e^{C_n(X,Y/2)}\right)^\dagger \e^{Y/2}\e^{X}\e^{Y/2} \left(\prod_{n=2}^\infty \e^{C_n(X,Y/2)}\right)\nonumber  \\
  =& C_R \e^X C_R^\dagger.\nonumber
\end{align*}
In the first line we have applied the Zassenhaus formula. In the second line we have used that $X$, $Y$, and $\exp(X+Y)$ are Hermitian. In the third line we have used that $Y/2$ commutes with itself. In the fourth line we applied the Zassenhaus formula again. 

The operator $C_R$ must be chosen such that $\sigma_{MR}$ is positive, and the constraint $\tr_M \sigma_{MR} = \rho_R$ holds. This can be guaranteed by choosing
\begin{align}
  C_R = \rho_R^{1/2} Z_R^{-1/2},
\end{align}
where
\begin{align}
  Z_R^{-1/2} =& \left\{ \tr_M \left[ \exp\left( \log \rho_M - \frac{1}{\alpha} H_{MR} \right) \right] \right\}^{-1/2}.
\end{align}

We thus have the following self-consistent equation for an optimal solution $\sigma_{MR}$:
\begin{align}\label{eq:app_A:sol2}
  \begin{aligned}
  \sigma_{MR} =& \rho_R^{1/2} Z_R^{-1/2} \exp\Bigg( \log \sigma_M - \frac{1}{\alpha} H_{MR}\Bigg)\\
  &\times Z_R^{-1/2}\rho_R^{1/2},
  \end{aligned}
\end{align}
The assosicated conditional state, given through \autoref{eq:app_A:condstate}, thus has the form claimed in \autoref{eq:sol1a}.

\subsection{\label{app:A:MX_CQ}Classical memories}

We now consider a situation where $M$ is restricted to be classical, while $X$ ($R$) and $Y$ are not. More precisely we consider the same optimization problem as before, but now restricted to optimizing over classical-quantum states of the form
\begin{align}\label{eq:appA:MX_CQ}
  \sigma_{MR} = \sum_m p(m) \ket{m}\bra{m} \otimes \sigma_{R|m},
\end{align}
for a fixed basis $\{\ket{m}\}$ on $M$ and arbitrary states $\sigma_{R|m}$ on $R$. The constraint $\tr_M \sigma_{MR} = \rho_R$ must still be satisfied. Note that the set of all such states is convex, when the basis for $M$ is fixed. 

The optimization proceeds as before, leading to \autoref{eq:app_A:sol2}, which when using \autoref{eq:appA:MX_CQ} gives us
\begin{align}
  \begin{aligned}
  p_m\sigma_{R|m} =&\, p_m\, \rho_R^{1/2} Z_R^{-1/2} \exp\left(\frac{1}{\alpha} \mathcal{R}^\dagger\left(\log \sigma_{Y|m} - \log \sigma_Y \right) \right) \\
  &\times Z_R^{-1/2} \rho_R^{1/2}, \\
  Z_R =& \sum_m p_m \exp\left(\frac{1}{\alpha} \mathcal{R}^\dagger\left(\log (\sigma_{Y|m}) - \log \sigma_Y \right) \right).
  \end{aligned}
\end{align}
Together with
\begin{align}
  \begin{aligned}
  &p_m = \tr[p_m\sigma_{R|m}],\\
  &\sigma_{Y|m} = \mathcal{R}(\sigma_{R|m}),
  \end{aligned}
\end{align}
this is a self-consistent equation for $p_m\sigma_{R|m}$.

\subsection{\label{app:A:fullyclassical}Fully Classical Setup}

We now consider a situation where all three systems are entirely classical in the sense that we can write
\begin{align}
  \rho_{R} =& \sum_{x} p(x) \ket{x}\bra{x},\\
  \rho_{MR} =& \sum_{m,x} p(m,x) \ket{m}\bra{m}\,\ket{x}\bra{x},\\
  \rho_{MY} =& \sum_{m,y} p(m,y) \ket{m}\bra{m}\,\ket{y}\bra{y},
\end{align}
and
\begin{align}\label{eq:appB:R_CC}
\mathcal{R}_{Y|R} = \sum_{x,y} p(y|x) \ket{y}\bra{y} \, \ket{x}\bra{x},
\end{align}
where $\ket{m}$, $\ket{x}$ and $\ket{y}$ are orthonormal bases for the three respective systems, $M$, $X$ and $Y$, $p(x)$, $p(m,x)$ and $p(m,y)$ are joint probability distributions, and $p(y|x)$ a conditional probability distribution. By plugging this into \autoref{eq:app_A:sol2}, it is straight forward to show that we can write
\begin{align}
  p(m,x) = \frac{p(m)}{p(x)} \exp \left(\frac{1}{\alpha} \sum_y p(y|x) \log\left(\frac{p(m,y)}{p(m) p(y)} \right) \right) Z_R^{-1},
\end{align}
where
\begin{align*}
  &p(m) = \sum_x p(m,x) \\
  &p(m,y) = \sum_x p(y|x) p(m,x).
\end{align*}
The normalization factor now reads
\begin{align}
  Z_R = \sum_m p(m) \exp \left( \frac{1}{\alpha} \sum_y p(y|x) \log\left(\frac{p(m,y)}{p(m) p(y)} \right) \right).
\end{align}
This is identical to the result for the classical information bottleneck, derived in \cite{Tishby99}.

\section{\label{app:B}Properties of an optimal solution}

\subsection{The negative of $I_\text{pred}$ as ``energy'' in the low $\alpha$ limit}

In \autoref{sect:temp} we noted that for small $\alpha$, the conditional quantum operator in \autoref{eq:sol1a} has a form analogous to a thermal state with $\alpha$ playing the role of temperature and $H_{MR}$ the role of Hamiltonian. We now show that the expectation value of this operator is just the negative of the predictive power:
\begin{align}
  \begin{aligned}
  \braket{H_{MR}} &:= \tr_{MR} [\sigma_{MR} H_{MR}] \\
  =&\,-\tr_{MR} [\sigma_{MR} \hat{\mathcal{R}}^\dagger\left(\log\sigma_{MY} - \log(\sigma_{M}\otimes\sigma_Y) \right)] \\
  =&\,-\tr_{MY} [\hat{\mathcal{R}}(\sigma_{MR}) \left(\log\sigma_{MY} - \log(\sigma_{M}\otimes\sigma_Y) \right)] \\
  =&\,-\tr_{MY} [\sigma_{MY} \left(\log\sigma_{MY} - \log(\sigma_{Y}\otimes\sigma_Y) \right)] \\
  =&\,-I(\sigma_{MY})\\
  =&\,-I_\text{pred},
  \end{aligned}
\end{align}
In the second line we inserted the definition of $H_{MR}$, in the third we used the defining property of the dual map, $\hat{\mathcal{R}}^\dagger$, in the fourth that $\sigma_{MY} = \hat{\mathcal{R}}(\sigma_{MR})$, and in the last two lines we simply used the definitions of the mutual information and $I_\text{pred}$, respectively.

\subsection{No quantum advantage for classical processes} \label{nqa}

In this section we prove that there is no quantum advantage for predictive inference on classical processes, as claimed in \autoref{sect:noquadvantage}.

Recall that the relevance channel, $\mathcal{R}^{R\to Y}$, is said to be \emph{classical} if it can be written in terms of a conditional quantum operator of the form
\begin{align}\label{eq:appB:R_CC}
\mathcal{R}_{Y|R} = \sum_{x,y} p(y|x) \ket{y}\bra{y} \, \ket{x}\bra{x},
\end{align}
where $p(y|x)$ is a conditional probability distribution, and $\ket{x}$ and $\ket{y}$ are orthonormal basis sets for the two systems $R$ and $Y$ (recall also that $R$ is isomoprhic to $X$).

We first show that applying a map of this form to an arbitrary state $\sigma_{MR}$ results in a quantum-classical state on $MY$, \emph{i.e.}, a state of the form
\begin{align}\label{eq:appB:MY_QC}
\sigma_{MY} = \hat{\mathcal{R}}(\sigma_{MR}) = \sum_y p(y) \sigma_{M|y} \ket{y}\bra{y},
\end{align}
where $p(y)$ is a probability distribution and $\sigma_{M|y}$ are quantum states on $M$:
\begin{align}
\begin{aligned}
    \hat{\mathcal{R}}(\sigma_{MR}) =& \tr_R\left[\sum_{x,y}p(y|x) \ket{y}\bra{y}\,\ket{x}\bra{x} \sigma_{MR}\right]\\
    =& \sum_{x,y} p(y|x) \braket{x|\sigma_{MR}|x}\, \ket{y}\bra{y}\\
    =& \sum_y p(y) \sigma_{M|y}\ket{y}\bra{y},
\end{aligned}
\end{align}
where
\begin{align}
    p(y) \sigma_{M|y} = \sum_x p(y|x) \braket{x|\sigma_{MR}|x}.
\end{align}

We next show that if the initial state, $\rho_R$, is diagonal in the chosen basis $\ket{x}$,
\begin{align}\label{eq:appB:X_diag}
\rho_R = \sum_x p(x)\ket{x}\bra{x},
\end{align}
then \autoref{eq:sol1a} implies that the optimal state $\rho_{MR}^\text{opt}$ corresponding to an optimal encoding is also quantum-classical. For ease of notation we drop the superscript ``opt'' for the rest of this section.

Recall that we showed in \app{app:A} that the state $\rho_{MR}$ corresponding to an optimal encoding must satisfy
\begin{align}
  \begin{aligned}
  \rho_{MR} =& \rho_R^{1/2} Z_R^{-1/2} \exp\Bigg( \log \rho_M - \frac{1}{\alpha} H_{MR}\Bigg)\\
  &\times Z_R^{-1/2}\rho_R^{1/2},
  \end{aligned}
\end{align}
where
\begin{align}
  H_{MR} =& \hat{\mathcal{R}}^\dagger\left(\log\rho_{M} + \log\rho_Y - \log\rho_{MY}\right),\\
  Z_R^{-1/2} =& \left\{ \tr_M \left[ \exp\left( \log \rho_M - \frac{1}{\alpha} H_{MR} \right) \right] \right\}^{-1/2}.
\end{align}
Using that $\rho_{MY}$ is quantum-classical we can write
\begin{align}
\begin{aligned}
\log \rho_{MY} =& \log\left[\sum_y p(y) \rho_{M|y}\ket{y}\bra{y}\right]\\
=& \sum_y \log\left[p(y) \rho_{M|y}\right] \ket{y}\bra{y}\\
=& \sum_y \left[\log p(y) + \log\rho_{M|y}\right] \ket{y}\bra{y}\\
=& \log \rho_Y + \sum_y \log\left(\rho_{M|y}\right) \ket{y}\bra{y}.
\end{aligned}
\end{align}
Thus,
\begin{align}
\begin{aligned}
H_{MR} =& \hat{\mathcal{R}}^\dagger\left[\log \rho_M - \sum_y \log\left(\rho_{M|y}\right) \ket{y}\bra{y}\right]\\
=& \log \rho_M - \sum_{x,y} p(y|x) \log\left(\rho_{M|y}\right) \ket{x}\bra{x}\\
=& \sum_x \left[ \log \rho_M - \sum_{y} p(y|x) \log\left(\rho_{M|y}\right) \right] \ket{x}\bra{x}
\end{aligned}
\end{align}
where we have used that $\mathcal{R}^\dagger(I_Y) = I_R = \sum_x \ket{x}\bra{x}$ and
\begin{align}
\mathcal{R}^\dagger\left(\ket{y}\bra{y}\right) = \sum_x p(y|x) \ket{x}\bra{x}.
\end{align}
Using this, we have that
\begin{align}
\begin{aligned}
&\exp\Bigg( \log \rho_M - \frac{1}{\alpha} H_{MR}\Bigg) \\
&= \sum_x \exp\Bigg( \left(1-\frac{1}{\alpha}\right)\log \rho_M  \\
&\,\,\,\,\,\,\,\,\,\,\,\,\,\,\,\,\,\,\,\,\,\,\,\,\,\,\,\,\,\,\,\,\,+ \frac{1}{\alpha} \sum_{y} p(y|x) \log\left(\rho_{M|y}\right) \Bigg) \ket{x}\bra{x}\\
&= \sum_x \tau_{M|x} \ket{x}\bra{x},
\end{aligned}
\end{align}
where we have defined
\begin{align}\label{eq:appB:tau_Mgivenx}
\begin{aligned}
\tau_{M|x} =& \exp\Bigg( \left(1-\frac{1}{\alpha}\right)\log \rho_M \\
&\,\,\,\,\,\,\,\,\,\,\,\,\,\,\,\,+ \frac{1}{\alpha} \sum_{y} p(y|x) \log\left(\rho_{M|y}\right) \Bigg).
\end{aligned}
\end{align}
This means that we can write
\begin{align}
  \begin{aligned}
  \rho_{MR} =& \sum_x \tau_{M|x} \rho_R^{1/2} Z_R^{-1/2} \ket{x}\bra{x} Z_R^{-1/2}\rho_R^{1/2},
  \end{aligned}
\end{align}
and
\begin{align}
  \begin{aligned}
  Z_R = \sum_x \tr_M[\tau_{M|x}]\ket{x}\bra{x}.
  \end{aligned}
\end{align}
Now, it just remains to use that $\rho_R = \sum_x p(x) \ket{x}\bra{x}$ to arrive at
\begin{align}\label{eq:appB:MX_QC}
  \begin{aligned}
  \rho_{MR} =& \sum_x p(x) \rho_{M|x} \ket{x}\bra{x},
  \end{aligned}
\end{align}
where
\begin{align}\label{eq:appB:rho_Mgivenx}
\rho_{M|x} = \frac{\tau_{M|x}}{\tr[\tau_{M|x}]}.
\end{align}
This proves the claim that $\rho_{MR}$ is quantum-classical.

The form of \autoref{eq:appB:MX_QC} does not rule out some degree of quantumness of the memory, since the states $\rho_{M|x}$ need not be orthogonal. However, we now show that this does not allow for higher predictive power at fixed memory, thus ruling out any quantum advantage.

First of all, note that the maximum possible value of $I_\text{pred}$ for a classical process can be reached by a state of the form
\begin{align}\label{eq:appB:MX_CC}
\rho_{MR} = \sum_{m,x} p(x)p(m|x) \ket{m}\bra{m}\,\ket{x}\bra{x},
\end{align}
which we refer to as a \emph{classical state}, since both $M$ and $R$ are classical. This can be achieved by choosing $M \cong R$ and $p(m|x) = \delta_{m,x}$, where $\delta_{m,x}$ is the Kronecker delta. It is easily seen that this gives the same output state $\rho_{MY}$ as would a purification $\psi_{XR}$ sent through the relevance channel. We therefore have that for any achievable predictive power there exists a classical state of the form of \autoref{eq:appB:MX_CC} that reaches this value.

Now, consider the memory and predictive power of the state \autoref{eq:appB:MX_QC}:
\begin{align}
I_\text{mem}(\rho_{MR}) =& S(\rho_M) - \sum_x p(x) S(\rho_{M|x}),\\
I_\text{pred}(\rho_{MR}) =& S(\rho_M) - \sum_y p(y) S(\rho_{M|y}),
\end{align}
where the output state after sending through the relevance channel is
\begin{align}
\rho_{MY} = \hat{\mathcal{R}}(\rho_{MR}) = \sum_{y} p(y)\rho_{M|y}\ket{y}\bra{y},
\end{align}
and $p(y)\rho_{M|y} = \sum_x p(y|x) p(x) \rho_{M|x}$. Using Eqs. \eqref{eq:appB:rho_Mgivenx} and \eqref{eq:appB:tau_Mgivenx} we find that the Lagrangian, $L(\rho_{MR})$, that we wish to maximize is
\begin{align}
  \begin{aligned}
  L(\rho_{MR}) =& I_\text{pred}(\rho_{MR}) - \alpha I_\text{mem}(\rho_{MR}) \\
  =&  \sum_x p(x) \log\left(\tr\,\tau_{M|x}\right).
  \end{aligned}
\end{align}
Since the right hand side only depends on the trace of $\tau_{M|x}$, the Lagrangian is invariant under a measurement of the memory $M$ in a chosen basis $\ket{m}$. This shows that quantum correlations of the type where $\rho_{M|x}$ are non-orthogonal are superfluous. We conclude that optimal values of the Lagrangian can be reached for classical states in the case of classical processes.

\section{\label{sect:algo}An iterative algorithm and deterministic annealing}

In this appendix we introduce an algorithm to find optimal encodings for given initial data and relevance channel. This is a direct quantum generalization of the Information Bottleneck method~\cite{Tishby99} used to find such optimal encodings in the classical case. The Information Bottleneck method employs an iterative algorithm where a solution at one iteration is used on the right hand side of \autoref{eq:sol1a} to compute a solution for the next iteration. This algorithm is similar to the Blahut-Arimoto algorithm, for which each iteration is proven to converge~\cite{Cover12,Blahut72}, but since the functional we are optimizing is not convex over the product space of states on $MR$ and $MY$ uniqueness is not guaranteed~\cite{Tishby99}. The Information Bottleneck method has been used in several domains, and it has inspired new algorithms as well as new insights into existing data analysis methods which can be derived from the elegant conceptual framework (see e.g. \cite{friedman_multivariate_2001, dhillon_information-theoretic_2003, StillBialek2004, StillBialBot04, GIB, bekkerman_multi-way_2005, Still-IAL09, creutzigB, StillPrecup12, Still2014, jacoby_information_2015, tishby_deep_2015}, and references therein).

We introduce the following quantum generalization to find an optimal encoding at a given value of $\alpha$. Denoting the state at the $k$'th step by $\sigma_{MR}^{(k)}$, the state at the next step is found through
\begin{widetext}
\begin{enumerate}
  \item Compute $\sigma_M^{(k)}$, $\sigma_{MY}^{(k)}$ and $\sigma_Y^{(k)}$ according to
  \begin{align*}
    \sigma_M^{(k)} = \tr_R \sigma_{MR}^{(k)}, \qquad
    \sigma_{MY}^{(k)} = \hat{\mathcal{R}}(\sigma_{MR}^{(k)}),\qquad
    \sigma_Y^{(k)} = \tr_M \sigma_{MY}^{(k)}.
  \end{align*}
  \item Compute the state for the next step according to:
  \begin{align*}
    \sigma_{MR}^{(k+1)} =&\, \rho_R^{1/2} \left(Z_R^{(k)}\right)^{-1/2} 
  \exp\Bigg( \log \sigma_M^{(k)} - \frac{1}{\alpha} H_{MR}^{(k)}\Bigg) 
  \left(Z_R^{(k)}\right)^{-1/2} \rho_R^{1/2},
  \end{align*}
  where
  \begin{align*}
  H_{MR}^{(k)} = \hat{\mathcal{R}}^\dagger\left(\log(\sigma_{M}^{(k)} + \log\sigma_Y^{(k)} - \log\sigma_{MY}^{(k)}\right),
  \end{align*}
  and
  \begin{align*}
    Z_R^{(k)} =& \tr_M \bigg[ \exp\Big( \log \sigma_M^{(k)} - \frac{1}{\alpha} H_{MR}^{(k)}  \Big) \bigg].
  \end{align*}
\item Halt if $L^{(k+1)}-L^{(k)} \le \varepsilon$, for some tolerance $\varepsilon$, where $L^{(k)} = I_\text{pred}\left(\sigma_{MR}^{(k)}\right)-\alpha I_\text{mem}\left(\sigma_{MR}^{(k)}\right)$.
\end{enumerate}

Whenever we restrict to a situation where the memory is classical, as discussed in \app{app:A:MX_CQ}, we replace step 3 with
\begin{enumerate}
\setcounter{enumi}{2}
  \item Compute the state for a classical memory for the next step according to:
\begin{align*}
  p_m^{(k+1)}\sigma_{R|m}^{(k+1)} =& p_m^{(k)} \rho_R^{1/2} (Z_R^{(k)})^{-1/2} \times \exp\left(-\frac{1}{\alpha} \hat{\mathcal{R}}^\dagger\left(\log \sigma_Y^{(k)} - \log \sigma_{Y|m}^{(k)}\right) \right) \times (Z_R^{(k)})^{-1/2} \rho_R^{1/2},
\end{align*}
where
\end{enumerate}
\begin{align*}
  Z_R^{(k)} =& \sum_m p_m^{(k)} \exp\left(-\frac{1}{\alpha} \hat{\mathcal{R}}^\dagger\left(\log \sigma_Y^{(k)} - \log (\sigma_{Y|m}^{(k)}) \right) \right).
\end{align*}
And step 2 is replaced with $p_m^{(k)} = \tr\left[p_m^{(k)}\sigma_{A|m}^{(k)}\right]$, and $\sigma_{B|m}^{(k)} = \mathcal{E}(\sigma_{A|m}^{(k)})$. This allows us to compare the optimal solution found for a classical memory to that of a quantum memory.
\end{widetext}

The iterative algorithm allows us to find a solution at given $\alpha$, where $\alpha$ parametrizes the tradeoff between $I_\text{pred}$ and $I_\text{mem}$. By varying this parameter, we can trace out a curve in the ``information plane''~\cite{Tishby99,Still10} with $I_\text{mem}$ on the x-axis and $I_\text{pred}$ on the y-axis. This is analogous to a rate-distortion curve in conventional rate-distortion theory. To trace out this curve we use the following scheme: we start at a large value of $\alpha$ with the trivial solution $\rho_{MR}^\text{opt} = \rho_M\otimes\rho_R$ ($\rho_M$ arbitrary), and gradually ``cool'' the system by lowering $\alpha$ towards zero in small steps. At each step we use a small perturbation of the solution at the previous step as initial guess. This approach to clustering problems has been dubbed ``deterministic annealing,'' due to the analogy with statistical mechanics~\cite{Rose98}. 

The annealing scheme allows us not only to find the optimal solution at a fixed memory dimension, $d_M$, but also to find the smallest $d_M$ possible at each value of $\alpha$. At very large $\alpha$ we have the trivial solution $\rho_M \otimes \rho_R$ for arbitrary $\rho_M$, and we can therefore choose $M$ to be a trivial system with Hilbert space dimension $d_M = 1$. As we lower $\alpha$ we compare the solution for a memory of dimension $d_M$ to one with dimension $d_M+1$, and increase the dimension for the next step, $d_M \leftarrow d_M+1$, only if the higher dimensional system outperforms the lower dimensional one. In this way, the optimal solution will go through a series of ``phase transitions''~\cite{Rose98} at critical values of $\alpha$, where the dimension is increased.

It is particularly interesting that in the $\alpha\to 0$ limit of this annealing scheme one obtains a solution with maximal predictive power equal to that of a purification, but potentially at a lower $I_\text{mem}$ and memory dimension $d_M$.

\bibliography{refs}

\end{document}